%% file: arxiv.tex
\PassOptionsToPackage{unicode}{hyperref}
\PassOptionsToPackage{hyphens}{url}
\PassOptionsToPackage{dvipsnames,svgnames,x11names}{xcolor}
\documentclass[12pt]{article}

\usepackage{amsmath,amssymb}
\usepackage{iftex}
\ifPDFTeX
  \usepackage[T1]{fontenc}
  \usepackage[utf8]{inputenc}
  \usepackage{textcomp} 
\else 
  \usepackage{unicode-math}
  \defaultfontfeatures{Scale=MatchLowercase}
  \defaultfontfeatures[\rmfamily]{Ligatures=TeX,Scale=1}
\fi
\usepackage{lmodern}
\ifPDFTeX\else  
\fi
\IfFileExists{upquote.sty}{\usepackage{upquote}}{}
\IfFileExists{microtype.sty}{
  \usepackage[]{microtype}
  \UseMicrotypeSet[protrusion]{basicmath} 
}{}
\makeatletter
\@ifundefined{KOMAClassName}{
  \IfFileExists{parskip.sty}{%
    \usepackage{parskip}
  }{
    \setlength{\parindent}{0pt}
    \setlength{\parskip}{6pt plus 2pt minus 1pt}}
}{
  \KOMAoptions{parskip=half}}
\makeatother
\usepackage{xcolor}
\makeatletter
\ifx\paragraph\undefined\else
  \let\oldparagraph\paragraph
  \renewcommand{\paragraph}{
    \@ifstar
      \xxxParagraphStar
      \xxxParagraphNoStar
  }
  \newcommand{\xxxParagraphStar}[1]{\oldparagraph*{#1}\mbox{}}
  \newcommand{\xxxParagraphNoStar}[1]{\oldparagraph{#1}\mbox{}}
\fi
\ifx\subparagraph\undefined\else
  \let\oldsubparagraph\subparagraph
  \renewcommand{\subparagraph}{
    \@ifstar
      \xxxSubParagraphStar
      \xxxSubParagraphNoStar
  }
  \newcommand{\xxxSubParagraphStar}[1]{\oldsubparagraph*{#1}\mbox{}}
  \newcommand{\xxxSubParagraphNoStar}[1]{\oldsubparagraph{#1}\mbox{}}
\fi
\makeatother

\usepackage{longtable,booktabs,array}
\usepackage{calc} 
\usepackage{etoolbox}
\makeatletter
\patchcmd\longtable{\par}{\if@noskipsec\mbox{}\fi\par}{}{}
\makeatother
\IfFileExists{footnotehyper.sty}{\usepackage{footnotehyper}}{\usepackage{footnote}}
\makesavenoteenv{longtable}
\usepackage{graphicx}
\makeatletter
\def\maxwidth{\ifdim\Gin@nat@width>\linewidth\linewidth\else\Gin@nat@width\fi}
\def\maxheight{\ifdim\Gin@nat@height>\textheight\textheight\else\Gin@nat@height\fi}
\makeatother
\setkeys{Gin}{width=\maxwidth,height=\maxheight,keepaspectratio}
\makeatletter
\def\fps@figure{htbp}
\makeatother

\makeatletter
\@ifpackageloaded{caption}{}{\usepackage{caption}}
\AtBeginDocument{%
\ifdefined\contentsname
  \renewcommand*\contentsname{Table of contents}
\else
  \newcommand\contentsname{Table of contents}
\fi
\ifdefined\listfigurename
  \renewcommand*\listfigurename{List of Figures}
\else
  \newcommand\listfigurename{List of Figures}
\fi
\ifdefined\listtablename
  \renewcommand*\listtablename{List of Tables}
\else
  \newcommand\listtablename{List of Tables}
\fi
\ifdefined\figurename
  \renewcommand*\figurename{Figure}
\else
  \newcommand\figurename{Figure}
\fi
\ifdefined\tablename
  \renewcommand*\tablename{Table}
\else
  \newcommand\tablename{Table}
\fi
}
\@ifpackageloaded{float}{}{\usepackage{float}}
\floatstyle{ruled}
\@ifundefined{c@chapter}{\newfloat{codelisting}{h}{lop}}{\newfloat{codelisting}{h}{lop}[chapter]}
\floatname{codelisting}{Listing}

\makeatother
\makeatletter
\@ifpackageloaded{caption}{}{\usepackage{caption}}
\@ifpackageloaded{subcaption}{}{\usepackage{subcaption}}
\makeatother

\ifLuaTeX
  \usepackage{selnolig}  
\fi
\usepackage[]{natbib}
\usepackage{bookmark}

\usepackage[noblocks]{authblk}

\IfFileExists{xurl.sty}{\usepackage{xurl}}{} 
\hypersetup{
  pdftitle={GeoMix: A Bayesian Framework for Stratigraphic and Geotechnical Inference},
  pdfauthor={Bradley Wakefield; Michael Bertolacci; David Gunawan; Fraser Bransby; Phil Watson; Andrew Zammit-Mangion},
  pdfkeywords={Bayesian hierarchical modelling; Markov random field; Gaussian process; Vecchia approximation; Offshore site characterisation},
  colorlinks=true,
  linkcolor={blue},
  filecolor={Maroon},
  citecolor={Blue},
  urlcolor={Blue},
  pdfcreator={LaTeX via pandoc}}

\usepackage{amsthm,amsfonts,refcount,placeins,makecell,setspace}
\usepackage{algorithm}
\usepackage{algorithmicx}
\usepackage{algpseudocode}
\usepackage{tikz}
\usetikzlibrary{matrix,fit,positioning,calc,arrows.meta}
\usepackage{tabularx,ragged2e}
\usepackage{rotating}
\usepackage{caption}

\newcolumntype{Y}{>{\RaggedRight\arraybackslash}X}
\newcolumntype{P}[1]{>{\RaggedRight\arraybackslash}p{#1}}

\newcommand{\svec}{\mathbf{s}}
\newcommand{\xvec}{\mathbf{x}}

\newcommand{\Zvec}{\mathbf{Z}}
\newcommand{\Yvec}{\mathbf{Y}}
\newcommand{\Xvec}{\mathbf{X}}
\newcommand{\Epvec}{\boldsymbol{\varepsilon}}
\newcommand{\Gs}{\mathcal{G}}
\newcommand{\Hs}{\mathcal{H}}
\newcommand{\Cs}{\mathcal{C}}
\newcommand{\p}{\mathcal{M}}
\newcommand{\thetab}{\boldsymbol{\theta}}
\newcommand{\Sigmab}{\boldsymbol{\Sigma}}

\newcommand{\alphab}{\boldsymbol{\alpha}}

\newcommand{\mub}{\boldsymbol{\mu}}

\newcommand{\Ss}{\mathcal{S}}

\newcommand{\anon}{1}

\begin{document}

\def\spacingset#1{\renewcommand{\baselinestretch}{#1}\small\normalsize} \spacingset{1}


\if1\anon
{
  \title{GeoMix: A Bayesian Framework for Stratigraphic and Geotechnical Inference}
 \author[1]{Bradley Wakefield}
\author[2]{Michael Bertolacci}
\author[1]{David Gunawan}
\author[3]{Fraser Bransby}
\author[3]{Phil Watson}
\author[4,1]{Andrew Zammit-Mangion}

\affil[1]{School of Mathematics and Physics,
University of Wollongong, NSW 2522, Australia}

\affil[2]{Department of Mathematics and Statistics,
University of Western Australia, WA 6009, Australia}

\affil[3]{Centre for Offshore Foundation Systems,
School of Earth and Oceans,
University of Western Australia, WA 6009, Australia}

\affil[4]{School of Mathematics and Statistics,
University of New South Wales, NSW 2052, Australia}
  \maketitle
} \fi

\if0\anon
{
  \bigskip
  \bigskip
  \bigskip
  \begin{center}
    {\LARGE\bf GeoMix: A Bayesian Framework for Stratigraphic and Geotechnical Inference}
\end{center}
  \medskip
} \fi

\begin{abstract}
Accurate characterisation of subsea sediments is essential for engineering projects such as offshore wind farms, but the challenging environment and large spatial domains involved make dense geotechnical measurements impractical. Seismic traces offer a cheap, abundant complement, yet their geotechnical interpretation relies on subjective judgement, leading to biased, hard-to-validate predictions of sediment structure.
To address this, we introduce GeoMix, a fully Bayesian hierarchical framework for jointly inferring geological stratigraphy and geotechnical properties from seismic and geotechnical data. GeoMix uses a Potts Markov random field to represent spatially coherent geology organised into distinct stratigraphic layers. Labels derived from seismic interpretation are linked to the latent stratigraphy through an observation model that captures elevated uncertainty near transitions between strata. Conditional on the latent stratigraphy, geotechnical properties are modelled using class-specific Gaussian processes, with scalable likelihood evaluation enabled by a grouped Vecchia approximation.
Together, these components yield a unified latent-variable model capable of recovering stratigraphy, quantifying uncertainty, and predicting geotechnical properties across large three-dimensional domains. Simulation studies and an application to the IJmuiden Ver offshore wind farm show that GeoMix produces spatially coherent stratigraphic estimates and sediment predictions that are better calibrated and more accurate than existing approaches.
\end{abstract}

\noindent%
{\it Keywords:} Bayesian hierarchical modelling; Markov random field; Gaussian process; Vecchia approximation; Offshore site characterisation
\vfill

\spacingset{1.9} 
\captionsetup{
  font={stretch=1}
}

\section{Introduction}
Accurate characterisation of the mechanical properties of the seabed is essential for the safe and efficient design of offshore infrastructure. Limited understanding of sediment properties and stratigraphy complicates foundation design, constructability, and exposure to geotechnical hazards \citep{Randolph2017}, often leading in practice to conservative over-design that can hinder the delivery of large offshore projects \citep{Baecher2023}. This problem is most acute for offshore wind farms, where individual projects can span hundreds of square kilometres of seabed and include numerous foundation or anchor locations. Seabed data collection is limited to either sparse but relatively accurate (geotechnical) data or abundant but indirect (geophysical) data. The need to reduce costs and maintain economic viability at these scales motivates approaches that can synthesise heterogeneous data sources while quantifying uncertainty in both sediment properties and geological layering.

Offshore site characterisation draws on complementary but incomplete geotechnical and geophysical data sources. Geotechnical data such as borehole sampling with laboratory testing provide precise point measurements of sediment properties, but are expensive, time-consuming, and spatially sparse across large offshore developments \citep{Randolph2017}. In situ tests, particularly cone penetration tests (CPTs), provide high vertical resolution measurements of seabed properties, though they also remain sparse at the horizontal scale of offshore wind farm sites \citep{lunne2011,Robertson2016}. CPT data are widely used to delineate sediment layering and infer seabed properties, primarily through transformations of cone tip resistance, $q_c$, sometimes alongside sleeve friction and pore pressure measurements. Many foundation design methods use depth profiles of cone tip resistance directly as design inputs (e.g., the methods for axial pile-capacity design in \citealp{API2014}). Consequently, recent site investigation practice for offshore wind farms has placed significant emphasis on achieving $q_c$ profiles over the entire site, informed by relatively sparse CPT results \citep{Peuchen2022}. 

Geophysical investigations, including seismic and acoustic surveys, provide broad spatial coverage that informs the mapping of the boundaries between geological strata \citep{Tsai2022}. These methods infer sediment properties indirectly and rely on subjective interpretation that must be calibrated against limited geotechnical ground‑truth data \citep{Dawson2005}. Reliability and precision can also degrade at greater depths, and in highly variable seabeds \citep{Baecher2023}. In practice, geotechnical and geophysical data are synthesised into a \emph{ground model} that represents stratigraphy and stratum-specific properties, interpolates between sparse investigations, and provide key inputs to foundation design and geotechnical risk management (e.g. \citealp{Rattley2017}; \citealp{Thomas2017}).

The modelling and prediction of geotechnical properties at a site present substantial challenges \citep{Phoon2022}. Such properties typically exhibit strong vertical nonstationarity, pronounced horizontal variability in both material properties and layer depths, and marked anisotropy, with horizontal spatial correlations generally exceeding vertical correlations \citep{Phoon1999,Oguz2020,Ching2021,geowarp}. These features have motivated many statistical approaches for spatial prediction which do not incorporate ground model information, including classical geostatistical methods \citep{Degroot1993,Vessia2020}, random field formulations \citep{Fenton1999,Uzielli2005}, and Bayesian compressive sampling \citep{Wang2017b}, supported more broadly by probabilistic and Bayesian frameworks for uncertainty propagation and parameter estimation in geotechnical design \citep{Lacasse2007}. 

Gaussian process (GP) models provide a coherent framework for spatial prediction and uncertainty quantification and have been applied in offshore settings to represent the strong spatial dependence observed in CPT data \citep{Cai2019,Sastre2021}. To accommodate nonstationarity, \citet{geowarp} proposed warped Gaussian process models (\emph{GeoWarp}), which infer spatially varying dependence through adaptive transformations learned from CPT observations. However, stratigraphic organisation and class-dependent sediment behaviour can induce abrupt transitions and highly nonstationary dependence structures that remain difficult to capture with many geostatistical approaches, often leading to oversmoothing and poor representation of structural heterogeneity \citep{Ray2019,Oguz2020}. In particular, such approaches represent stratigraphy only implicitly and do not provide an explicit probabilistic representation of seabed layering.

A natural alternative is to model stratigraphy as a latent categorical field coupled to geotechnical observations. Hidden Markov random field (MRF) formulations represent this as a spatial mixture model, in which a spatially coherent process governs class membership and class-conditional observation models link these labels to geotechnical properties \citep{Li2016,Wang2017,Bolin2019,Shuku2023}. Prior geological information can constrain the latent field to more plausible configurations; \citet{Li2016} conditioned the MRF prior on borehole-derived stacking sequences, while \citet{Bolin2019} demonstrated that latent Gaussian Markov random field mixture (LGFM) models can integrate fine-scale information to infer permeability fields at larger spatial scales. Despite their appeal, applying MRF-based spatial mixture models in realistic offshore domains raises several practical challenges. Mixture models can be sensitive to initialisation and may exhibit label switching when not anchored by observed labels \citep{Li2016}. Scalable methods that explicitly represent stratigraphy and are robust to imperfect prior stratigraphic information are therefore needed.

In this paper, we propose \emph{GeoMix}, a Bayesian spatial mixture framework for offshore site characterisation that jointly models stratigraphy via a Markov random field and geotechnical properties via Gaussian processes. In contrast to existing approaches, GeoMix incorporates an existing ground model as a noisy proxy for stratigraphy and provides a probabilistic mechanism that accommodates misclassification, missing and misaligned boundaries, and class-dependent CPT behaviour, while remaining computationally scalable over large three-dimensional grids. The framework enables joint inference on geotechnical properties and stratigraphic boundaries with calibrated uncertainty.

\section{The IJmuiden Ver wind farm zone}
\label{sec:IJV}

\begin{figure}[t]
    \centering
    \includegraphics[]{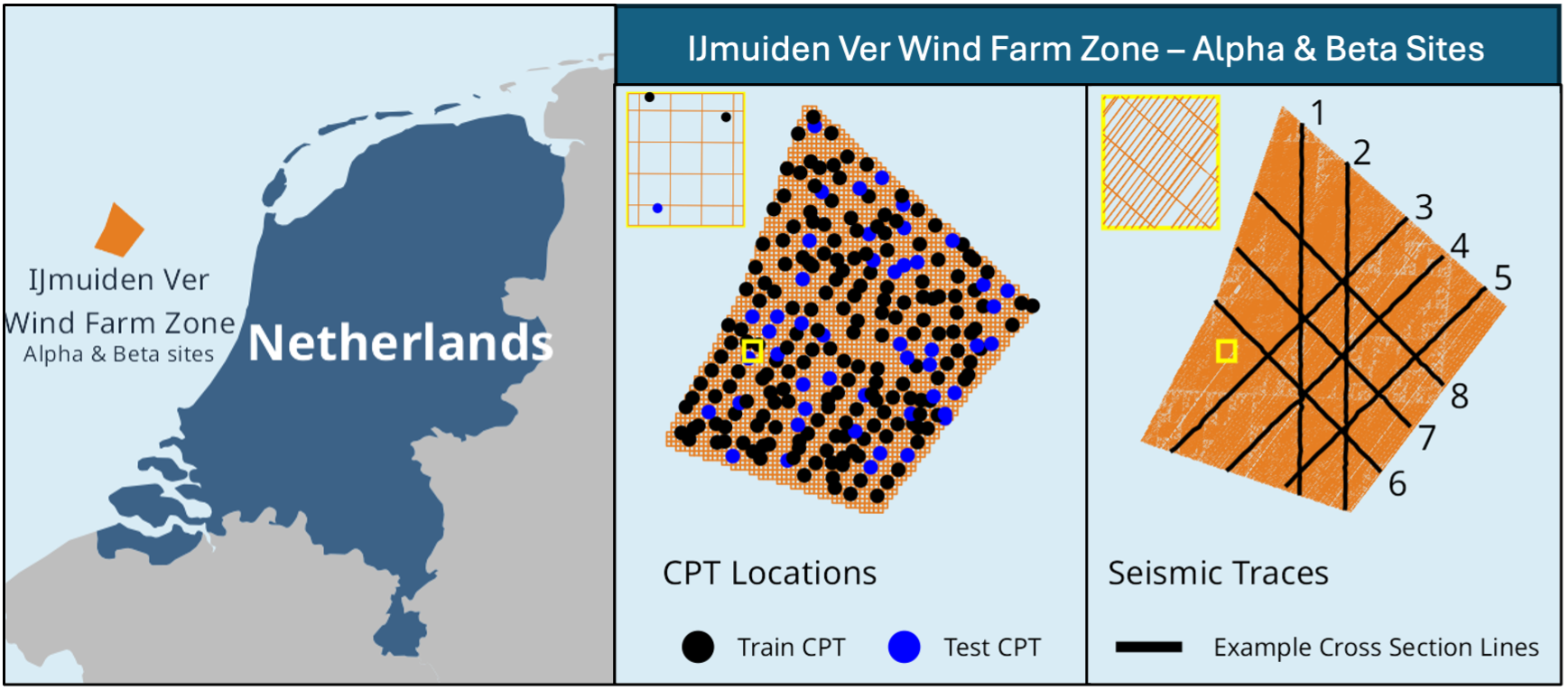}
    \caption{Location of the IJmuiden Ver offshore wind farm zone in the Netherlands and the North Sea (left panel). The middle panel shows cone penetration test (CPT) locations used for model fitting (black) and testing (blue) over the horizontal spatial lattice on which our GeoMix model is constructed. The right panel shows the spatial footprint of the seismic traces used to construct the ground model, with black lines indicating the eight cross-section transects used for visualisation (see Section~\ref{sec:app-results}).}
    \label{fig:map}
\end{figure}

The IJmuiden Ver offshore wind farm zone is situated approximately 54\,km northwest of the Dutch coastline in the North Sea (Figure \ref{fig:map}, left panel). It forms part of the Netherlands’ national offshore renewable energy program, established by the Dutch Government to expand large-scale wind energy capacity. The concession area encompasses roughly 400\,km$^2$ and, upon full development, is projected to deliver up to 6\,GW of installed generation capacity. The zone is subdivided into four designated lease sites (Alpha, Beta, Gamma A, and Gamma B), each delineated for phased construction and geotechnical investigation. In this study, we restrict attention to the Alpha and Beta sites. We refer to these sites collectively as the IJV-AB site. 

The IJV-AB site presents a challenging subsea characterisation problem. The seabed comprises multiple sedimentary units that vary laterally and vertically. The available observations are heterogeneous: high coverage and noisy seismic interpretations provide categorical indicators of stratigraphy, while sparse CPTs offer precise but spatially limited measurements of geotechnical properties. Reconciling these complementary data sources into a coherent, uncertainty-quantified model of the underlying geology requires methodology capable of handling spatial dependence, layer boundaries, and class-specific continuous geotechnical data within a hierarchical modelling framework. The central aim of this paper is to develop a model that leverages both seismic and geotechnical information to produce probabilistic stratigraphic maps and predictive distributions of engineering parameters (specifically, cone tip resistance) across the IJV-AB site. Further details are available in the technical documentation provided by the Netherlands Enterprise Agency \citep{IJV2023}.

\subsection{Geophysical site investigation}

Extensive geophysical surveying has been conducted across the IJV-AB site. Seismic trace data were acquired along 437 survey lines at roughly 7\,m intervals (see Figure \ref{fig:map} right panel). The seismic signals penetrate to depths of up to approximately 80\,m below the seabed surface, although data quality typically degrades with increasing depth due to attenuation and scattering effects. 

Seismic measurements are derived from wave amplitudes obtained through 2D ultra-high-resolution multi-channel seismic (2D-UHRS) reflection surveys, which capture contrasts in acoustic impedance between sediment layers and are converted from two-way travel time to depth using an interval-velocity-based depth–time model. The resulting impedance profiles provide indirect information on subsea stratigraphy and are further processed to reduce noise and enhance reflection continuity. The processed sections are then reviewed by geophysical interpreters, who delineate stratigraphic boundaries and assign preliminary layer information based on the amplitude and continuity of acoustic impedance contrasts, informed where available by supporting geotechnical data. Ultimately, these interpreted profiles are synthesised into a \emph{ground model}, described in Section \ref{sec:ground_model}.

\subsection{Geotechnical site investigation}
Complementing the geophysical surveys, a comprehensive geotechnical investigation was undertaken using CPTs and boreholes. A total of 238 CPTs were performed across the IJV-AB site, providing high-resolution vertical profiles of cone tip resistance (and also sleeve friction and $u_2$ pore pressure). These CPT profiles typically extend from the seabed to depths of between 40\,m and 80\,m below the seafloor and provide direct measurements of soil behaviour that are essential for calibration and validation of the seismic-derived ground model discussed next. The CPT data exhibit pronounced vertical heterogeneity, reflecting rapid transitions between coarse and fine-grained sediment layers, as well as strong lateral variability associated with deviations in the geological stratification and composition. 

\subsection{Ground model}
\label{sec:ground_model}
\begin{figure}[t]
    \centering
    \includegraphics[]{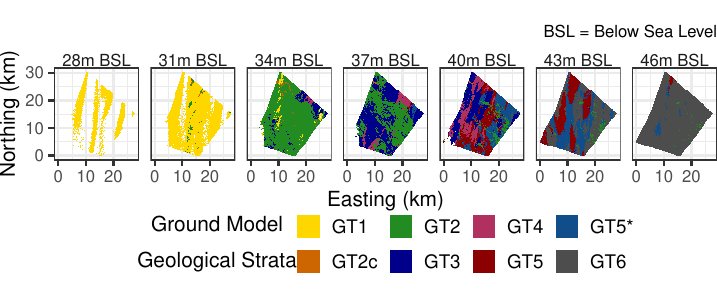}
    \caption{Ground model interpreted geological strata across horizontal slices of the IJV-AB domain. The eight units shown represent the identified geological layers used throughout the analysis, each corresponding to a distinct sedimentary layer (see Table~\ref{supp:tab:geounits} in the Supplementary Material for details).}
    \label{fig:app_grid_z1}
\end{figure}

Integration of seismic and geotechnical information forms the basis of the ground model used throughout this study. Geological strata (‘units’) were delineated by interpreters at a coarser lateral resolution than the raw seismic sampling, with an effective horizontal spacing of approximately 70\,m.  Interpretations of CPT profiles together with laboratory testing of borehole samples were then used to assign sediment classifications and representative geotechnical properties to each unit. The resulting geological strata (denoted GT1, GT2, GT2c, GT3, GT4, GT5, GT5*, and GT6) represent the identified layers at the site, each corresponding to a distinct sedimentary unit with characteristic material properties and depositional history (see Table~\ref{supp:tab:geounits} in the Supplementary Material for a detailed description). Figure~\ref{fig:app_grid_z1} visualises the resulting ground model via horizontal slices at 3\,m depth increments, highlighting the large-scale stratigraphic architecture of the IJV--AB site. Lateral discontinuities, abrupt transitions, and channelised or patchy features reflect both genuine geological variability and uncertainty inherent in UHRS-based interpretation.

\begin{figure}[ht]
    \centering
    \includegraphics[]{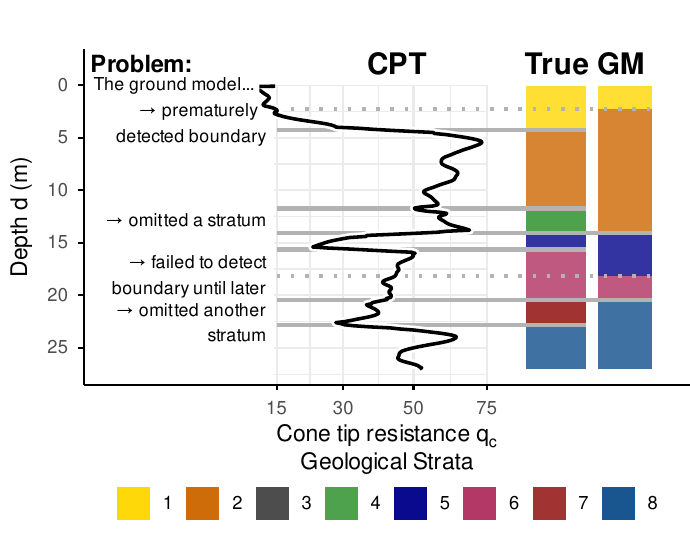}
    \caption{Stylised illustration of inconsistencies between ground model (GM) stratigraphic boundaries and the underlying 'true' stratigraphy, and their relationship to CPT-derived cone tip resistance $q_c$. Shaded columns indicate stratum assignments as a function of depth for the latent stratigraphy and the ground model, while the solid line shows the CPT cone tip resistance on the Box–Cox transformed scale ($\lambda = 0.6$). Solid horizontal lines denote true stratigraphic interfaces associated with marked changes in the $q_c$ profile, whereas dotted lines indicate misaligned boundaries in the ground model.}
    \label{fig:misspec}
\end{figure}

Although the ground model provides a site-wide delineation of stratigraphic boundaries, these boundaries remain uncertain in areas of complex geology. Moreover, they are sometimes inconsistent with the geotechnical measurements from the CPTs. As illustrated in a stylised manner in Figure~\ref{fig:misspec}, the ground model can (i) introduce boundaries where the CPT-derived cone tip resistance ($q_c$) shows no meaningful change, (ii) omit strata transitions evident in CPT data, or (iii) place interfaces above or below depths suggested by geotechnical measurements. These discrepancies arise from limitations in depth--time conversion and artefacts introduced during seismic processing and interpretation. Consequently, the ground model provides a noisy and potentially biased proxy for the underlying stratigraphy. These discrepancies motivate the probabilistic framework developed in this paper, which treats the ground model as a noisy observational proxy, and quantifies uncertainty in both stratigraphy and geotechnical properties.

\section{Spatial statistical model}
\label{sec:model}
Consider spatial locations indexed by vectors $\svec \equiv (x, y, d)'$, where $x$ and $y$ denote lateral coordinates and $d$ denotes depth.  
Let $Y_1(\cdot)$ denote a latent categorical field, defined over $\Ss \subset \mathbb{R}^3$, representing the underlying geological stratigraphy, which assigns to each location $\svec \in \mathcal{S}$ a class $Y_1(\svec) \in \{1, \ldots, K\}$.   
Similarly, let $Y_2(\cdot) \in \mathbb{R}$ denote a continuous-valued latent process defined over $\mathcal{S}$, representing a latent geotechnical property (throughout, we consider only a transformed version of cone tip resistance, $q_c$).  

Let $\mathcal{S}_1 \equiv \{ \svec_{1,1}, \ldots, \svec_{1,N_1} \}$ denote a regular three-dimensional spatial lattice of $\mathcal{S}$ of size $N_1$, comprising the set of locations at which the latent geological strata are modelled. Assume further that each grid location is associated with a noisy geological stratum label from the ground model.
We denote by $\Zvec_1 \equiv (Z_{1,1}, \ldots, Z_{1,N_1})'$ the collection of labels derived from the ground model at these locations, where each $Z_{1,i} \in \{1, \ldots, K\}$ is a noisy representation of $Y_1(\svec_{1,i})$, for $i = 1,\dots,N_1$.  
We use the notation $\Yvec_1 \equiv (Y_{1,1}, \ldots, Y_{1,N_1})'$ to denote the classes of the actual geological strata, where $Y_{1,i} \equiv Y_1(\svec_{1,i})$, for $i = 1,\ldots,N_1$.  

Let $\mathcal{S}_2 \equiv \{ \svec_{2,1}, \ldots, \svec_{2,N_2} \} \subseteq \mathcal{S}_1$ denote the subset of locations at which continuous geotechnical measurements are collected via CPTs (in practice, $N_2 \ll N_1$). 
We write $\Zvec_2 \equiv (Z_{2,1}, \ldots, Z_{2,N_2})'$, where each $Z_{2,i} \in \mathbb{R}$ is a noisy measurement of $Y_2(\svec_{2,i})$, for $i = 1,\dots,N_2$.  
We similarly write $\Yvec_2 \equiv (Y_{2,1}, \ldots, Y_{2,N_2})'$, where each $Y_{2,i} \equiv Y_2(\svec_{2,i})$ denotes the corresponding latent process, for $i = 1,\dots,N_2$. The observed geotechnical ($q_c$) measurements of $\Yvec_2$ are treated as noisy realisations of this latent process, with additive measurement error,
\begin{equation}
\Zvec_2 = \Yvec_2 + \Epvec, \quad \text{where} \quad \Epvec \equiv (\varepsilon_1, \ldots, \varepsilon_{N_2})',\quad \varepsilon_i \overset{\text{iid}}{\sim} \mathcal{N}(0, \tau^2).
\label{eq:Z2}
\end{equation}
Here, $\Epvec$ collects independent Gaussian noise terms, and $\tau^2 > 0$ denotes the measurement-error variance, which we assume is constant for all locations in $\mathcal{S}_2$. 

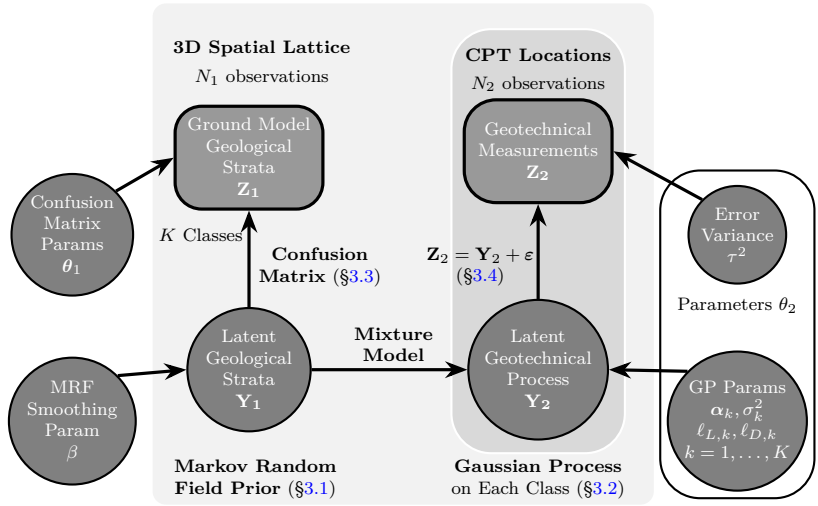
\begin{figure}[t]
\centering
\begin{tikzpicture}[
    >=Stealth,
    scale=0.85,
    transform shape,
    font=\scriptsize,
    node distance=2.5cm,
    latent/.style={circle,draw=black,fill=gray!100,
                   minimum size=1.55cm,thick,text=white,align=center,
                   inner sep=1pt},
    param/.style={circle,draw=black,fill=gray!100,
                  minimum size=1.15cm,thick,text=white,align=center,
                  inner sep=1pt},
    obsZone/.style={rounded corners=15pt,draw=white,
                    fill=gray!30,thick,minimum width=2.7cm,
                    minimum height=6.6cm,align=center},
    theta1Zone/.style={rounded corners=15pt,draw=black,
                       fill=white,thick,minimum width=2.4cm,
                       minimum height=3.1cm,align=center},
    obsZoneInner/.style={rounded corners=12pt,draw=black,
                         fill=gray!80,very thick,
                         minimum width=2.3cm,minimum height=1.6cm,
                         align=center,text=white,inner sep=2pt},
    obsZbox/.style={rounded corners=10pt,draw=black,
                    fill=gray!80,very thick,minimum width=2.3cm,
                    minimum height=1.6cm,align=center,text=white,
                    inner sep=2pt},
    labeltxt/.style={align=center,inner sep=1pt}
]

\spacingset{1}

\fill[gray!10,rounded corners=5pt] (1.5,8.7) rectangle (9.3,0.9);
\node[obsZone] (Z2outer) at (7.5,5) {};
\fill[theta1Zone] (9.4,6.1) rectangle (11.8,1);

\node[latent] (Y1) at (3,3) {Latent\\Geological\\Strata\\$\mathbf{Y_1}$};
\node[latent] (Y2) at (7.5,3) {Latent\\Geotechnical\\Process\\$\mathbf{Y_2}$};

\node[param] (theta2) at (10.6,2.2)
  {GP Params\\$\alphab_k,\sigma^2_k$\\$\ell_{L,k},\ell_{D,k}$\\$k=1,\ldots,K$};

\node[param] (theta1) at (0.25,5.1)
  {Confusion\\Matrix\\Params\\$\thetab_1$};

\node[param] (beta) at (0.25,2.2)
  {MRF\\Smoothing\\Param\\$\beta$};

\node[param] (tau) at (10.6,5.1)
  {Error\\Variance\\$\tau^2$};

\node[obsZbox] (Z1) at (3,6.3)
  {Ground Model\\Geological\\Strata\\$\mathbf{Z_1}$};

\node[labeltxt] at (7.5,7.9) {\textbf{CPT Locations}};
\node[labeltxt] at (7.5,7.45) {$N_2$ observations};
\node[obsZoneInner] (Z2inner) at (7.5,6.4)
  {Geotechnical\\Measurements\\$\mathbf{Z_2}$};

\draw[very thick,->] (Y1.north) -- (Z1.south);
\draw[very thick,->] (Y2.north) -- (Z2inner.south);
\draw[very thick,->] (Y1.east) -- (Y2.west);
\draw[very thick,->] (theta1.north east) -- (Z1.west);
\draw[very thick,->] (theta2.north west) -- (Y2.east);
\draw[very thick,->] (beta.north east) -- (Y1.west);
\draw[very thick,->] (tau.north west) -- (Z2inner.east);

\node[labeltxt] at (5.25,3.4) {\textbf{Mixture}\\\textbf{Model}};
\node[labeltxt] at (2.25,5.1) {$K$ Classes};
\node[labeltxt,anchor=east] at (5.1,4.6)
  {\textbf{Confusion}\\\textbf{Matrix} (\textsection\ref{sec:model-Z1})};
\node[labeltxt] at (3.1,1.3)
  {\textbf{Markov Random}\\\textbf{Field Prior} (\textsection\ref{sec:model-Y1})};
\node[labeltxt] at (6.6,4.6)
  {$\Zvec_2 = \Yvec_2 + \Epvec$\\(\textsection\ref{sec:model-Z2})};
\node[labeltxt] at (7.5,1.3)
  {\textbf{Gaussian Process}\\on Each Class (\textsection\ref{sec:model-Y2})};

\node[labeltxt] at (3.2,8) {\textbf{3D Spatial Lattice}};
\node[labeltxt] at (3.2,7.55) {$N_1$ observations};
\node[labeltxt] at (10.6,4) {Parameters $\theta_2$};

\end{tikzpicture}

\caption{Graphical representation of the GeoMix hierarchical structure.}
\label{fig:modeldiagram}
\end{figure}

Figure \ref{fig:modeldiagram} provides a graphical overview of the model and the relationships between the variables, which are formalised in the remainder of this section.

\subsection{Latent spatial field of geological strata}
\label{sec:model-Y1}
We model the latent spatial field of geological strata $Y_1(\cdot)$ over $\Ss$ using a discrete lattice representation $\Yvec_1$. Specifically, for $\svec \in \Ss$ we write $Y_1(\svec) =  \mathbf{1}(\svec)'\Yvec_1$, where $\mathbf{1}(\svec)$ is an indicator vector whose $i$th element, for $i=1,\ldots,N_1$, equals $1$ if $\svec$ is closest to lattice site $\svec_{1,i}$ and $0$ otherwise.

To encode spatial coherence of geological strata while allowing sharp transitions at boundaries between strata, we place a Potts-style Markov random field prior on $\Yvec_1$. Namely, for each $Y_{1,i}$ corresponding to location $\svec_{1,i} \in \Ss_1$, conditioned on a smoothing parameter $\beta$ and all other points in $\Ss_1$, (i.e., $\Yvec_{1}^{\setminus i} \equiv \big(Y_1(\svec_1):\svec_1 \in \Ss_1 \setminus \{\svec_{1,i}\}\big)),$
the full conditional distribution is
\begin{equation}
\Pr(Y_{1,i} = k \mid \beta, \Yvec_{1}^{\setminus i})\ \propto\ 
\exp\!\left\{-\beta\ \sum_{j\sim i}  f\big(Y_{1,j},k\big)\right\},
\label{eq:potts}
\end{equation}
where $j\sim i$ denotes adjacency on the grid, and $f(\cdot,\cdot)$ is a measure of class disagreement. In this paper, we use $f(a,b)=1$ for $a\neq b$ and $f(a,b)=0$ otherwise, unless stated otherwise. The parameter $\beta\geq 0$ controls the overall strength of spatial smoothing.

Although the location-specific conditional distribution \eqref{eq:potts} is sufficient for location-wise Gibbs updates of the latent geological strata, the joint density of $\Yvec_1$ is computationally intractable due to a normalisation constant requiring summation over all possible configurations of $\Yvec_1$. We therefore adopt the following pseudo-likelihood approximation to estimate $\beta$,
\begin{equation}
\tilde p(\Yvec_1 \mid \beta)
\ \equiv\
\prod_{\svec_{1,i}\in \Ss_1}
\Pr\!\left(Y_{1,i} \mid \beta, \Yvec_{1}^{\setminus i}\right),
\label{eq:pseudopotts}
\end{equation}
where each conditional term is given by \eqref{eq:potts}.  This corresponds to the pseudo-likelihood method proposed by \citet{Besag1975}, which provides a tractable surrogate to the intractable Potts likelihood. For computational convenience, we estimate the Potts interaction parameter $\beta$ prior to model fitting (Section~\ref{sec:fit}) using \eqref{eq:pseudopotts}, with the ground model $\Zvec_1$ replacing the latent field $\Yvec_1$. We subsequently fix $\beta$ to the pseudo-likelihood-based estimate $\hat{\beta}$ throughout model fitting.
\subsection{Mixture distribution of the latent geotechnical property}
\label{sec:model-Y2}
We model the spatial process governing the latent geotechnical property of interest, $Y_2(\cdot)$, as a mixture of $K$ class-conditional Gaussian processes, each associated with a distinct geological stratum. To incorporate covariate effects into this specification, we construct a design matrix $\Xvec \in \mathbb{R}^{N_2 \times p}$ by stacking row vectors $\xvec(\svec_2)' \in \mathbb{R}^{1 \times p}$, where $\xvec(\svec_2)$ denotes the $p$-dimensional vector of covariate values at location $\svec_2 \in \mathcal{S}_2$.

For each class $k=1,\ldots,K$, define $\mathcal{S}_2^{(k)} \equiv \{\svec_2 \in \mathcal{S}_2 : Y_1(\svec_2)=k\}$, $n_k \equiv |\mathcal{S}_2^{(k)}|$, $\Yvec_2^{(k)} \equiv \left(Y_{2,i} : \svec_{2,i} \in \mathcal{S}_2^{(k)}\right)'$, $\Zvec_2^{(k)} \equiv \left(Z_{2,i} : \svec_{2,i} \in \mathcal{S}_2^{(k)}\right)'$, and $\Xvec^{(k)} \equiv \left(\xvec(\svec_{2,i}) : \svec_{2,i} \in \mathcal{S}_2^{(k)}\right)' \in \mathbb{R}^{n_k \times p}$.

We model the conditional distribution of $\Yvec_2$ to be block-independent and Gaussian,
\begin{equation}
  p\!\left(\Yvec_2 \mid \Yvec_1,\Xvec,\thetab_2\right)
  \;=\;
  \prod_{k=1}^K 
  p\Big(\Yvec_2^{(k)} \,\Big|\, \mub^{(k)},\ \tilde\Sigmab^{(k)}\Big),
  \qquad
  \Yvec_2^{(k)} \sim\mathcal{N}\left(\mub^{(k)},\ \tilde\Sigmab^{(k)}\right), 
  \label{eq:Y2}
\end{equation}
where $ \mub^{(k)} \equiv \Xvec^{(k)}\alphab_k$, $\alphab_k \in \mathbb{R}^p$ are class-specific regression coefficients, $\tilde{\Sigmab}^{(k)}$ denotes the covariance matrix for class $k$, and $\thetab_2$ is a collection of model parameters, detailed below.

The form of $\tilde{\Sigmab}^{(k)}$ is induced by a Matérn covariance function with smoothness parameter $\nu = {3}/{2}$. Specifically, for any pair of locations $\svec_{2,i_1} \equiv (x_{i_1}, y_{i_1}, d_{i_1})'$ and $\svec_{2,i_2} \equiv (x_{i_2}, y_{i_2}, d_{i_2})'$  sharing latent class $k$, the covariance between $Y_{2,i_1}$ and $Y_{2,i_2}$ is given by 
$
\mathrm{Cov}\!\big(Y_{2,i_1}, Y_{2,i_2}\big)
= \sigma_k^{2}
\big(1 + \sqrt{3}\,\Delta_{i_1,i_2}\big)
\exp\!\big(-\sqrt{3}\,\Delta_{i_1,i_2}\big),
$
where $\sigma_k^{2}$ is the marginal variance for class $k$, and  
the scaled anisotropic distance between the two locations is defined as
$
\Delta_{i_1,i_2}
= \sqrt{
  \frac{(x_{i_1} - x_{i_2})^2 + (y_{i_1} - y_{i_2})^2}{\ell_{L,k}^{2}}
  + \frac{(d_{i_1} - d_{i_2})^2}{\ell_{D,k}^{2}}
},
$
where $\ell_{L,k}$ and $\ell_{D,k}$ are the lateral and vertical length-scale parameters for class $k$, respectively, governing the rate of spatial correlation decay in each dimension. We collect the parameters across all classes together with the measurement error variance $\tau^2$ in $\thetab_2 \equiv \left((\alphab_k',\, \sigma_k^2,\, \ell_{L,k},\, \ell_{D,k})_{k = 1,\ldots,K},\tau^2\right)'$.

\subsection{Observed discrete labels of sedimentary layers}
\label{sec:model-Z1}
\begin{sidewaysfigure}
    \centering
    \includegraphics[width=\linewidth]{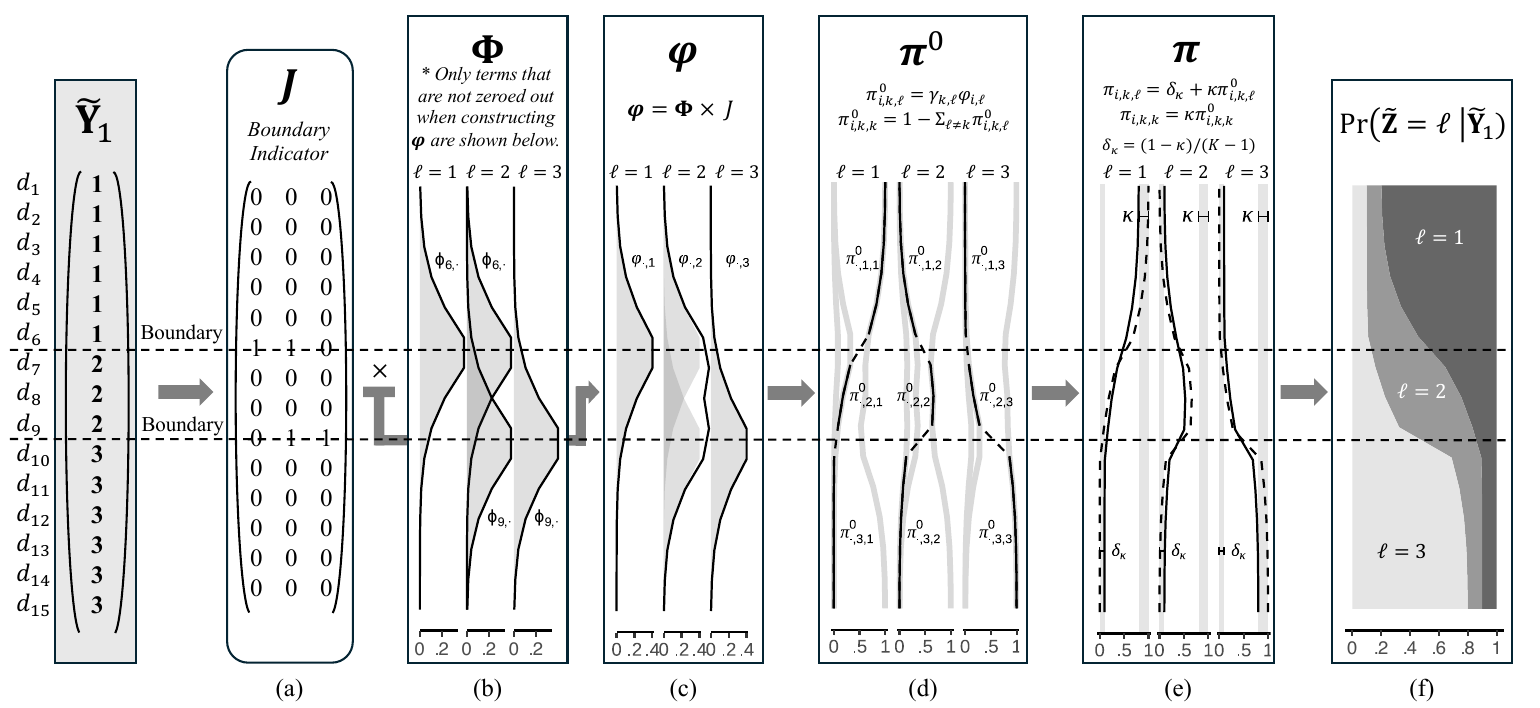}
    \caption{Construction of the boundary-aware confusion probabilities linking a vertical profile of the latent stratigraphy, $\tilde\Yvec_1$, and observed ground model labels, $\tilde\Zvec_1$, for an example depth profile with $w=15$ coordinates and $K=3$. (a) Boundary indicators $J_{r,\ell}$. (b) Local influence weights $\phi_{r,i}$. (c) Aggregated boundary influence $\varphi_{i,\ell}$. (d) Baseline misclassification probabilities $\pi^0_{i,k,\ell}$. (e) Renormalised probabilities $\pi_{i,k,\ell}$ with baseline $\delta_\kappa$. (f) Resulting categorical distribution $\Pr(\tilde Z_{1,i}=\ell \mid \tilde Y_1,\thetab_1)$.}
    \label{fig:confdiag}
\end{sidewaysfigure}

Seismic trace interpretation provides a noisy categorical label $\Zvec_1$ at all grid locations in the form of a ground model. The quantity $\Zvec_1$ differs from the latent categorical field $\Yvec_1$ because some strata are misclassified (recall Figure \ref{fig:misspec}). It is expected that most misclassifications occur near geological boundaries and diminish in probability with vertical distance from those interfaces. We therefore model misclassification through a boundary-aware confusion mechanism, computed within each vertical profile and dependent on the latent field $\Yvec_1$. 
The step-by-step construction is illustrated in Figure~\ref{fig:confdiag}. 

We construct a conditional model for depth-varying probabilities for the ground model ($\Zvec_1$) given the true underlying stratigraphy ($\Yvec_1$). We define the probability mass function of $\Zvec_1$ given $\Yvec_1$ and parameters $\thetab_1$ as
\begin{equation}
\Pr(\Zvec_1 \mid \Yvec_1, \thetab_1)
= \prod_{i=1}^{N_1} \Pr(Z_{1,i} \mid \Yvec_1, \thetab_1)
= \prod_{i=1}^{N_1} \pi_{i, Y_{1,i}, Z_{1,i}}.
\label{eq:Z1-likelihood}
\end{equation}
Here, $\pi_{i,Y_{1,i},\ell} \equiv \Pr(Z_{1,i} = \ell \mid \Yvec_1, \thetab_1)$ for $\ell \in \{1,\ldots,K\}$ denotes the contribution of a single observation $Z_{1,i}$. We detail the construction of $\pi_{i,Y_{1,i},\ell}$ below.

Consider a set of locations $\Omega \subseteq \Ss_1$ that contains points along a depth transect at a single horizontal location and, for ease of exposition, denote the depth coordinates in this transect as $d_1,...,d_w$, where $w = |\Omega|$. Denote the classes associated with this depth profile as $\tilde\Yvec_1$. Now, let $J_{r,\ell}$ indicate the presence of a class-$\ell$ interface at the midpoint between depths $d_r$ and $d_{r+1}$, for $r=1,\ldots,w-1$ (Figure~\ref{fig:confdiag}(a)). These indicators identify boundaries along each vertical profile that may induce confusion in the observed labels. To propagate the influence of such interfaces to nearby depths, we smooth the boundary indicators using a local kernel with possible unknown bandwidth $h>0$, where $\phi_{r,i} = \Phi\!\left(-\tfrac{1}{h}\left|\tfrac{d_{r+1}+d_r}{2}-d_i\right|\right)$, for $r=1,\ldots,w-1$ and $i=1,\ldots,w$, and $\Phi(\cdot)$ denotes the standard normal cumulative distribution function (Figure~\ref{fig:confdiag}(b)). The quantity $\phi_{r,i}$ determines how strongly an interface between depths $d_r$ and $d_{r+1}$ influences location $i$: the closer the interface, the stronger the influence. Aggregating these local contributions yields a boundary influence measure for each class, defined as $\varphi_{i,\ell} = \min\{\sum_{r}\phi_{r,i}J_{r,\ell},\,1\}$, for $i = 1,\ldots,w$ and $\ell = 1,\dots,K$, which quantifies the proximity of $d_i$ to class-$\ell$ interfaces within the vertical profile (Figure~\ref{fig:confdiag}(c)). Since $\phi_{r,i} \leq \Phi(0) = \tfrac{1}{2}$, a single boundary contributes at most $\tfrac{1}{2}$ to $\varphi_{i,\ell}$, preserving the distinction between proximity to one boundary and simultaneous proximity to two, as with a thin stratum where misclassification risk is greater.

Let $\gamma_{k,\ell}$ denote unknown weights controlling the degree of confusion between classes $k$ and $\ell$, with $\gamma_{k,\ell} \in (0,1)$ for $k \neq \ell$ and $\gamma_{k,k} = 0$ for all $k=1,\ldots,K$, such that $\sum_{\ell=1}^K\gamma_{k,\ell} \leq 1$. Using this parameter together with the boundary influence measure, we define so-called baseline probabilities for the ground model having class $\ell$ when the true class is $k$ as $\pi^0_{i,k,\ell} = \gamma_{k,\ell}\,\varphi_{i,\ell}$ for $k \neq \ell$, with $\pi^0_{i,k,k} = 1 - \sum_{\ell \neq k}\gamma_{k,\ell}\,\varphi_{i,\ell}$, so that $\sum_{\ell=1}^{K}\pi^0_{i,k,\ell} = 1$ (Figure~\ref{fig:confdiag}(d)). Non-negativity of $\pi^0_{i,k,k}$ is guaranteed by the constraint $\sum_\ell \gamma_{k,\ell} \leq 1$ together with $\varphi_{i,\ell} \leq 1$, which jointly ensure $\sum_{\ell\neq k}\gamma_{k,\ell}\,\varphi_{i,\ell} \leq 1$. Thus, $\pi^0_{i,k,\ell}$ depends on the true class $k$, the observed class $\ell$, and the proximity of location $i$ to interfaces involving class $\ell$.

The probabilities governing the ground model labels are obtained by renormalising these baseline quantities so that a non-zero baseline probability of misclassification, $\delta_\kappa$, is present at all depths. For $k = 1,\ldots,K$, we therefore set $\pi_{i,k,\ell} = \frac{1-\kappa}{K-1} + \kappa\,\pi^0_{i,k,\ell}$ if $k \neq \ell$, and $\pi_{i,k,k} = \kappa\,\pi^0_{i,k,k}$ otherwise, where $\kappa \in (0,1)$ controls the probability of correct classification and $\delta_\kappa = \frac{1-\kappa}{K-1}$ (Figure~\ref{fig:confdiag}(e)). This construction ensures that $\sum_{\ell=1}^{K}\pi_{i,k,\ell}=1$ and $\pi_{i,k,\ell}\geq\delta_\kappa$ for $k\neq\ell$. In practice we fix $\kappa$ close to one (we use $\kappa = 0.9$), reflecting the expectation that the ground model is predominantly correct within homogeneous regions away from boundaries. We verified that varying $\kappa$ over the range $[0.8, 0.95]$ did not materially alter posterior stratigraphic estimates or predictive performance in either the simulation or application setting. Following this construction we obtain $\pi_{i,Y_{1,i},\ell} \equiv \Pr(Z_{1,i} = \ell \mid \tilde\Yvec_1, \thetab_1)$ for $\ell = 1,\ldots,K$ and $i=1,\ldots,w$ (Figure~\ref{fig:confdiag}(f)).

We collect parameters associated with the misclassification mechanism in $\thetab_1 \equiv \left((\gamma_{k,\ell})_{k,\ell = 1,\ldots,K},\,h\right)'$. Further details on the construction and justification of this boundary-aware confusion model are provided in Section~\ref{supp:boundary-terms} of the Supplementary Material.

\subsection{Geotechnical measurements}
\label{sec:model-Z2}
From \eqref{eq:Z2} and \eqref{eq:Y2} 
\[
\Zvec_2^{(k)} \mid \Yvec_1,\Xvec,\thetab_2 
\sim \mathcal{N}\!\big(\mub^{(k)},\, \tilde{\Sigmab}^{(k)} + \tau^2 \mathbb{I}_{n_k}\big).
\]

Under our model, observations $\Zvec_2$ within each class are jointly correlated, leading to a likelihood involving large dense covariance matrices whose direct evaluation quickly becomes computationally prohibitive. To ensure computational tractability, we therefore adopt a grouped Vecchia approximation \citep{vecchia1988,Katzfuss2021}, described in detail in Section~\ref{supp:grouped-vecchia} of the Supplementary Material. The grouped Vecchia approximation expresses the likelihood function associated with each class as a product of low-dimensional conditional densities.  
This approximation leads to computational benefits, as updating the latent geological stratum at a given location requires evaluating a likelihood term involving only a small subset of $\Zvec_2$.

\subsection{Prior distributions}
\label{sec:prior}
We place prior distributions on all model parameters (except $\beta$, which is fixed to $\hat{\beta}$; see Section~\ref{sec:model-Y1}).
For each class $k=1,\ldots,K$, the regression coefficients follow
$\alphab_k \sim \mathcal{N}(\mathbf{0}_p, 100\,\mathbb{I}_p)$,
where $\mathbf{0}_p$ and $\mathbb{I}_p$ denote the $p$-dimensional zero vector and identity matrix, respectively.
The process and measurement-error variances are assigned inverse gamma priors,
$\sigma_k^2 \sim \mathrm{IG}(3,4)$ for $k=1,\ldots,K$ and
$\tau^2 \sim \mathrm{IG}(4,2)$, which concentrate prior mass moderate values consistent with the scale of Box--Cox-transformed cone tip resistance $q_c$.

To allow for anisotropic dependence, the lateral and vertical length-scale parameters are assigned exchangeable half-normal ($\mathrm{HN}(\cdot)$) priors,
$\ell_{L,k} \sim \mathrm{HN}(\sigma_{L}^2)$ and
$\ell_{D,k} \sim \mathrm{HN}(\sigma_{D}^2)$ for $k=1,\ldots,K$,
with scale hyperparameters assigned inverse-gamma priors
$\sigma_{L}^2 \sim \mathrm{IG}(3,5)$ and
$\sigma_{D}^2 \sim \mathrm{IG}(3,8)$.
Exchangeability allows information to be shared across classes, aiding estimation for narrow strata.

For the misclassification model,
$\boldsymbol{\eta}_k=(\eta_{k,1},\ldots,\eta_{k,K})
\sim \mathrm{Dirichlet}(\mathbf{1}_K)$ for $k=1,\ldots,K$,
where $\mathbf{1}_K$ is a length-$K$ vector of ones.
The basis weights are defined by $\gamma_{k,\ell}=\eta_{k,\ell}$ for $k\neq\ell$ and $\gamma_{k,k}=0$, which implies $\sum_{\ell=1}^K \gamma_{k,\ell} = 1-\eta_{k,k} \in (0,1)$ almost surely, satisfying the required constraint $\sum_{\ell=1}^K \gamma_{k,\ell} \leq 1$ and ensuring $\pi_{i,k,k} > 0$ almost surely (since $\varphi_{i,\ell}\leq 1$). The induced marginal prior on $(\gamma_{k,\ell})_{\ell\neq k}$ is a $(K-1)$-dimensional Dirichlet scaled by an independent $\mathrm{Beta}(K-1,1)$ factor. Importantly, this construction does not favour any class a priori and discourages sparse allocations. Although the $K(K-1)=56$ off-diagonal weights are numerous, their estimation is supported by the large number of vertical profiles and regularised by the Dirichlet prior; MCMC diagnostics confirmed good mixing for all $\gamma_{k,\ell}$ parameters across chains in both the simulation study and application.

Finally, the bandwidth parameter follows
$h \sim \mathrm{HN}(1)$, reflecting the expectation that misclassification is concentrated near stratum boundaries.

\subsection{Posterior inference via Markov chain Monte Carlo}
\label{sec:fit}

The joint posterior distribution over the latent field $\Yvec_1$, the confusion matrix parameters $\thetab_1$, and the Gaussian process parameters $\thetab_2$ does not admit a closed-form expression. Given observed data $\{\Zvec_1,\Zvec_2,\Xvec\}$ and the plug-in Potts parameter $\hat{\beta}$, the posterior is
\[
p(\Yvec_1,\thetab_1,\thetab_2 \mid \Zvec_1,\Zvec_2,\Xvec,\hat{\beta})
\propto
\underbrace{p(\Zvec_2 \mid \Yvec_1,\Xvec,\thetab_2)}_{\Zvec_2\text{ likelihood}}
\underbrace{p(\Zvec_1 \mid \Yvec_1,\thetab_1)}_{\Zvec_1\text{ likelihood}}
\underbrace{p(\Yvec_1 \mid \hat{\beta})}_{\text{Potts prior}}
\underbrace{p(\thetab_2)p(\thetab_1)}_{\text{parameter priors}}.
\]

We construct a Markov chain Monte Carlo (MCMC) scheme targeting this distribution. At each iteration, the sampler updates:
(i) the latent stratigraphic labels $\Yvec_1$ via Gibbs sampling;
(ii) the regression coefficients $\alphab_k$, for $k=1,\ldots,K$, via Gibbs updates exploiting conditional conjugacy;
(iii) the Gaussian process parameters $(\sigma_k^2, \ell_{L,k}, \ell_{D,k})$, $k=1,\ldots,K$, and the exchangeable prior parameters $\sigma_L^2$ and $\sigma_D^2$ using Hamiltonian Monte Carlo (HMC); 
(iv) the basis function weights $(\gamma_{k,\ell})$, $k,\ell=1,\ldots,K$, via Metropolis-within-Gibbs updates of $\boldsymbol{\eta}_k$, $k=1,\ldots,K$ (see Section~\ref{sec:prior}), using a Dirichlet proposal, and the confusion bandwidth parameter $h$ using HMC.
The full procedure is given in Algorithm~\ref{supp:alg:geomix-mcmc} of the Supplementary Material. All updates are implemented in \texttt{NIMBLE} \citep{nimble-software}. For step~(i), the sampler visits each location $\svec_{1,i} \in \mathcal{S}_1$ and samples $Y_{1,i}$ from its full conditional, factorising as
\begin{multline}
\label{eq:full_conditional_Y1}
\Pr\big(Y_{1,i}=k \mid \Yvec_{1}^{\setminus i}, \Zvec_1, \Zvec_2, \Xvec, \thetab_1, \thetab_2, \hat{\beta}\big) \propto {}\\
p\big(\Zvec_2 \mid Y_{1,i}=k, \Yvec_{1}^{\setminus i}, \Xvec, \thetab_2\big)\,
p\big(\Zvec_1 \mid Y_{1,i}=k, \Yvec_{1}^{\setminus i}, \thetab_1\big)\,
\Pr\big(Y_{1,i}=k \mid \Yvec_{1}^{\setminus i}, \hat{\beta}\big).
\end{multline}
Unnormalised log-conditionals are evaluated for each $k$, normalised to give a categorical distribution, and a new value is sampled. Details of each term and the prediction procedure are in Sections~\ref{supp:gibbs-detail} and \ref{supp:prediction} of the Supplementary Material.
\section{Simulation study}
\label{sec:sim}
We conducted a simulation study to assess out-of-sample prediction performance for both the latent geotechnical field $Y_2(\cdot)$ and latent stratigraphic field $Y_1(\cdot)$. 

\begin{figure}[ht]
    \centering
    \includegraphics[]{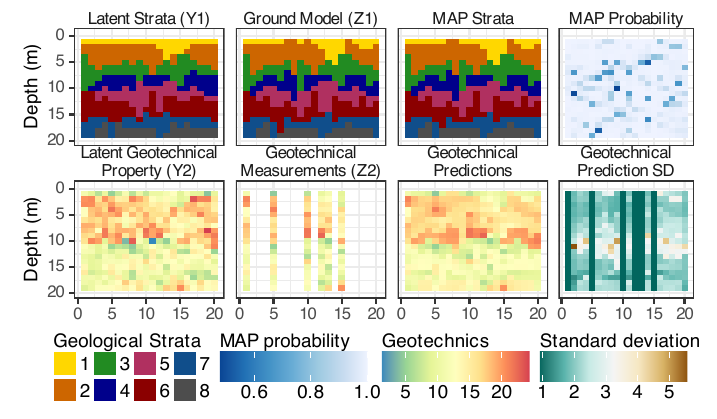}
    \caption{Cross-section through the synthetic domain for $y=9$. Top panels show (left to right) the latent geological strata, the ground-model strata, the MAP estimate of geological strata and the corresponding probability of the MAP estimate. The bottom panels show (left to right) the latent geotechnical property, the observed geotechnical measurements, the predicted geotechnical property obtained from GeoMix, and the standard deviation association with this prediction. 
}
    \label{fig:syn_cross}
\end{figure}

We generated data on a $20 \times 20 \times 20$ lattice ($N_1 = 8{,}000$ locations) with $K = 8$ geological strata and two covariates (intercept and depth). Continuous measurements were available at 160 randomly selected vertical profiles (40\% of the 400 profiles), with the remainder as prediction targets. Figure~\ref{fig:syn_cross} shows a cross section ($y=9$) of the simulated data including the latent strata and geotechnical property, the ground model, observed geotechnical data, and predictions from GeoMix.  

The latent stratigraphy was constructed by initialising vertically layered strata, sampling the Potts prior~\eqref{eq:potts} with $\beta=1.2$ and ordinal penalty $f(a,b)=|a-b|$ for 10 Gibbs sweeps, and enforcing vertical monotonicity within each column by iteratively correcting adjacent ordering violations. The latent geotechnical field was drawn from the class-conditional GP mixture of Section~\ref{sec:model-Y2} with Matérn ($\nu=3/2$) covariance, using class-specific parameters chosen to approximate Box--Cox-transformed ($\lambda=0.6$) cone tip resistance: $\alphab_0 = (10, 17, 14, 19, 7, 11, 6, 12)'$, $\alphab_1 = (0.20, 0.05, 0.20, 0.05, 0.10, 0.15, 0.20, 0.15)'$, $\boldsymbol{\sigma}^2 = (5, 4, 6, 5, 10, 4, 3, 7)'$, $\boldsymbol{\ell}_L = (0.5, 0.5, 0.6, 0.4, 0.6, 0.5, 0.7, 0.3)'$, $\boldsymbol{\ell}_D = (2.0, 1.8, 1.6, 1.7, 1.6, 2.0, 1.7, 1.5)'$, and $\tau^2 = 0.25$. Observations $\Zvec_2$ were generated via~\eqref{eq:Z2}. Ground-model labels $\Zvec_1$ were simulated using the misclassification model of Section~\ref{sec:model-Z1} with $h=2$, $\kappa=0.9$, and confusion weights
$\gamma_{k,\ell} = \tfrac{1}{2}\bigl(0.1 + \exp(-|k-\ell|)\bigr) \big/ \bigl(0.7 + \sum_{\ell \neq k}\exp(-|k-\ell|)\bigr)$
for $k\neq\ell$, with vertical monotonicity enforced. GeoMix was fitted using the same ordinal penalty $f(a,b)=|a-b|$ and the same fixed $\beta = 1.2$. We used depth-ordered Vecchia groups with conditioning sets of size 160 from the immediately preceding depth layer, with size chosen so that all observed vertical profiles are represented in each conditioning set (see Section~\ref{supp:grouped-vecchia} of the Supplementary Material for implementation details).

\subsection{Models}
\label{sec:sim-models}
We compared GeoMix against several competing approaches: a standard Gaussian process model (GP); the warped Gaussian process model GeoWarp (GW, \citealt{geowarp}); versions of both models incorporating ground model labels as covariates (GP$+\Zvec_1$ and GW$+\Zvec_1$); a linear model with ground model covariates (LM$+\Zvec_1$); and the latent Gaussian Markov random field mixture model (LGFM, \citealt{Bolin2019}), which uses a spatially dependent latent class field but does not incorporate $\Zvec_1$. Unlike the competing methods, GeoMix explicitly treats $\Zvec_1$ as a noisy observation of the latent geology, while LGFM classes are matched to geological strata post hoc using the Hungarian algorithm \citep{Kuhn1955,clue}.

\subsection{Simulated data results}
\label{sec:sim-results}
We evaluated each model on (i) prediction of $Y_2(\cdot)$ at $\mathcal{S}_1 \setminus \mathcal{S}_2$ and (ii) recovery of $\Yvec_1$. Performance metrics include root mean squared prediction error (RMSPE), coefficient of determination ($R^2$), bias, 95\% prediction interval coverage (Coverage), interval score (IS), continuous ranked probability score (CRPS), Dawid--Sebastiani score (DSS), a pairwise variant averaging the bivariate DSS over vertically adjacent pairs (DSS2), and logarithmic score (logS) \citep{crps,scoringrules,kruger2021,Gneiting2007,geowarp}. Convergence was assessed using four MCMC chains of $3{,}500$ post-burn-in iterations (following 500 burn-in iterations), thinned by $10$; diagnostics are provided in Section~\ref{supp:sim-mcmc} of the Supplementary Material.

\subsubsection{Predictive performance of the latent geotechnical property}
Table~\ref{tab:syn-metrics} reports averages across all performance metrics. GeoMix achieves the lowest RMSPE (8.309) and the most favourable scores across all proper scoring rule metrics (IS, CRPS, DSS, DSS2, logS). Among the models that do not explicitly use $\Zvec_1$, LGFM performs best overall, offering substantial improvements over the GP and GeoWarp models. However, LGFM is generally outperformed by models that incorporate $\Zvec_1$ information. Incorporating the ground model as a covariate yields consistent gains across the LM, GP, and GeoWarp models, underscoring the value of stratigraphic information (even when noisy) for predicting geotechnical properties. Although GW$+\Zvec_1$ improves upon its non-covariate counterparts, it remains outperformed by GeoMix on nearly all metrics, though GeoMix has slightly reduced coverage and higher bias. Note that although GeoMix exhibits the largest bias (0.090), all reported biases are negligible relative to the range of the true underlying $q_c$ ($-1.90$ to $27.69$). Predictions of $Y_2(\cdot)$ from GeoMix are shown in Figure~\ref{fig:syn_cross}.

\begin{table}[t]
\centering
\caption{Predictive performance metrics for each fitted model on the simulated dataset, predicting $Y_2(\cdot)$ evaluated on $\mathcal{S}_1 \setminus \mathcal{S}_2$. }
\label{tab:syn-metrics}
\renewcommand{\arraystretch}{0.8}
\begin{tabular}{lccccccccc}
\toprule
Model & RMSPE & Coverage & $R^2$ & Bias & IS & CRPS & DSS & DSS2 & logS \\
\midrule
GP     & 14.643 & 0.944 & 0.249 & 0.043 & 19.302 & 2.122 & 3.668 & 3.670 & 2.767 \\
GeoWarp     & 13.050 & 0.947 & 0.331 & 0.064 & 15.965 & 1.964 & 3.411 & 3.342 & 2.631 \\
LGFM & 11.111 & 0.944 & 0.430 & 0.050 & 15.132 & 1.767 & 3.289 & 3.274 & 2.500 \\
LM$+\Zvec_1$ & 9.474 & 0.941 & 0.514 & 0.028 & 17.255 & 1.653 & 3.259 & 3.260 & 2.664 \\
GP$+\Zvec_1$  & 9.474 & 0.941 & 0.514 & 0.028 & 17.200 & 1.651 & 3.255 & 3.256 & 2.654 \\
GW$+\Zvec_1$ & 9.089 & 0.935 & 0.534 & 0.034 & 15.830 & 1.622 & 3.134 & 3.095 & 2.516 \\
GeoMix & 8.309 & 0.928 & 0.574 & 0.090 & 13.364 & 1.513 & 2.952 & 2.948 & 2.349 \\
\bottomrule
\end{tabular}
\end{table}

\begin{figure}
    \centering
    \includegraphics[]{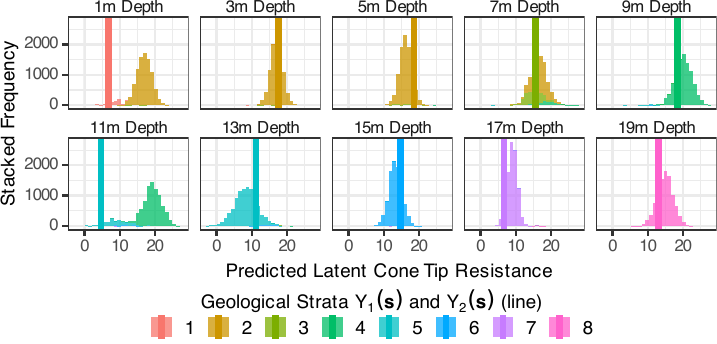}
    \caption{Predictive distributions of the latent cone tip resistance $Y_2(\cdot)$ for a test vertical profile at $(x, y) = (8, 7)$. Stacked histograms show predictive samples at each depth, with the colour of the histograms representative of the predicted $Y_1(\cdot)$ class, and with the true value marked as a vertical line whose colour indicates the actual $Y_1(\cdot)$ class.}
    \label{fig:syn_post_samp}
\end{figure}

Figure~\ref{fig:syn_post_samp} shows posterior samples of $Y_2(\cdot)$ for an example vertical profile at $(x,y) = (8,7)$. Within homogeneous strata, predictive distributions are approximately Gaussian and centred near the true values. Near geological boundaries, predictive distributions become multimodal as predictions switch between neighbouring classes. Even when boundary locations are imprecisely recovered, appreciable posterior mass falls on the correct configurations (see 1\,m, 7\,m, and 11\,m depths). 

\subsubsection{Predictive distribution of the latent class field}
We assess predictive performance for the latent stratigraphic field $\Yvec_1$ using both classification and probabilistic metrics. Table~\ref{tab:classification-metrics} reports accuracy, sensitivity, and specificity based on the most probable strata, along with the Brier score, cross-entropy, and AUC \citep{brier,yardstick} for $\Yvec_1$ across the observed and unobserved partitions ($\Ss_2$ and $\Ss_1 \setminus \Ss_2$), as well as across the full spatial domain ($\Ss_1$), with $\Zvec_1$ included as a baseline. GeoMix markedly outperforms LGFM across all partitions: AUC values of $0.991$--$0.997$ indicate near-perfect strata separation, and lower Brier and cross-entropy scores reflect sharper, better-calibrated posterior class probabilities. Relative to $\Zvec_1$, GeoMix attains comparable or higher accuracy while substantially improving probabilistic calibration. Accuracy gains are most pronounced at observed locations ($\Ss_2$); at unobserved locations the ground model label remains the MAP class in 99.2\% of cases, so point accuracy gains are modest, but GeoMix consistently provides improved uncertainty representation near class boundaries. The MAP strata are shown in Figure~\ref{fig:syn_cross}.

\begin{table}[t]
\centering
\caption{Classification performance metrics for $\Yvec_1$ for each model evaluated at geotechnical data locations ($\Ss_2$), unobserved locations ($\Ss_1 \setminus \Ss_2$), and across the entire lattice ($\Ss_1$).}
\label{tab:classification-metrics}
\begingroup
\setlength{\tabcolsep}{4pt}
\renewcommand{\arraystretch}{0.8}
\small
\begin{tabular}{llcccccc}
\toprule
\textbf{Subset} & \textbf{Model} & \textbf{Accuracy} & \textbf{Sensitivity} & \textbf{Specificity} & \textbf{Brier} & \textbf{Cross-Ent} & \textbf{AUC} \\
\midrule
\multicolumn{8}{l}{\footnotesize\emph{Note: Accuracy, sensitivity, and specificity computed under most-probable-class classification.}} \\
\hline
$\Ss_2$ & Z1     & 0.884 & 0.885 & 0.983 & 0.116 & -     & 0.934 \\
& LGFM   & 0.732 & 0.684 & 0.961 & 0.167 & 0.567 & 0.979 \\
& GeoMix & 0.942 & 0.945 & 0.992 & 0.046 & 0.213 & 0.997  \\ \hline
$\Ss_1 \setminus \Ss_2$ & Z1     & 0.870 & 0.869 & 0.981 & 0.130 & -     & 0.925 \\
& LGFM   & 0.617 & 0.577 & 0.945 & 0.237 & 0.794 & 0.958 \\
& GeoMix & 0.874 & 0.874 & 0.982 & 0.101 & 0.416 & 0.991 \\
\hline
$\Ss_1$ & Z1     & 0.876 & 0.875 & 0.982 & 0.124 & -     & 0.929 \\
& LGFM   & 0.663 & 0.619 & 0.951 & 0.209 & 0.703 & 0.967 \\
& GeoMix & 0.901 & 0.902 & 0.986 & 0.079 & 0.335 & 0.994 \\
\bottomrule
\end{tabular}
\endgroup
\end{table}

\section{IJmuiden Ver wind farm application}
\label{sec:app}

GeoMix was applied to the IJV-AB site by constructing a 3D lattice $\Ss_1$ with lateral spacing 350\,m $\times$ 350\,m and vertical spacing 0.5\,m. Depth coordinates were set at 0.5\,m intervals by uniformly downsampling the ground model. For lateral coordinates, the domain was partitioned into 350\,m $\times$ 350\,m cells, with each cell assigned the position of the CPT if a CPT falls within it, or the ground-model location closest to the cell centre otherwise. As a result, neighbouring locations in $\Ss_1$ need not be exactly equally spaced, although each point remains uniquely associated with a voxel in the underlying grid, yielding at most one CPT per voxel. The resulting lattice comprises $N_1 = 154{,}804$ locations, of which $|\mathcal{S}_2| = 7{,}665$ coincide with CPT observations used for model fitting and $2{,}009$ CPT observations held-out for testing. Cone tip resistance was transformed using a Box--Cox transformation with $\lambda = 0.6$,  supported by exploratory diagnostics, to reduce variance in the measurements and enhance the visibility of layer transitions; all models were fitted and evaluated on this scale. Vecchia groups were formed along the depth axis by merging depth levels with fewer than 160 observations; groups were ordered from shallow to deep, with each conditioning set of size 160 drawn from the immediately preceding depth group via a nearest-neighbour rule. The Potts parameter was estimated as $\hat{\beta} = 1.238$ via pseudo-likelihood on $\Zvec_1$. Owing to sparse lateral sampling relative to the spatial extent of the domain, lateral length-scales were weakly identifiable, with ranges smaller than the spacing between nearby CPT locations producing covariance structures that were difficult to distinguish from measurement noise and leading to poor mixing. The lateral length-scales were therefore fixed at their MAP estimates prior to MCMC, obtained via numerical optimisation of the joint posterior with the regression coefficients marginalised. Four MCMC chains were run in parallel for $3{,}500$ post--burn-in iterations (after 500 burn-in iterations), retaining every 10th sample. MCMC diagnostics are provided in Section~\ref{supp:ijv-mcmc} of the Supplementary Material; convergence is satisfactory across all parameters.

\subsection{Results}
\label{sec:app-results}
\begin{figure}[t]
    \centering
    \includegraphics[]{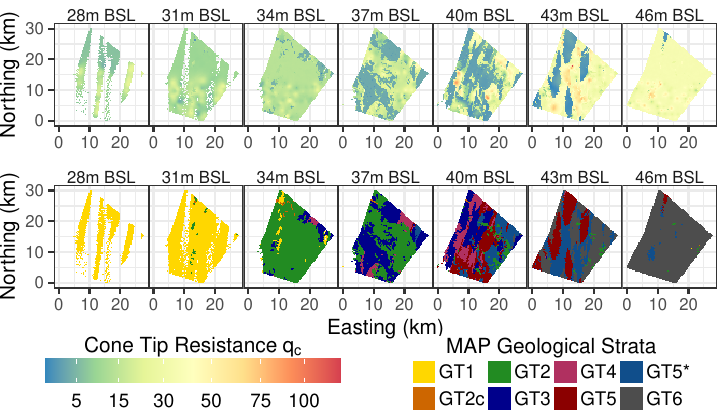}
    \caption{Posterior mean predictions of cone tip resistance $q_c$ (top) and the estimated latent field $Y_1(\cdot)$ across the IJV-AB site (bottom) from GeoMix. Values of $q_c$ on the Box–Cox ($\lambda = 0.6$) scale and the maximum-a-posteriori (MAP) geological stratum are shown.}
    \label{fig:app_grid_z2y1}
\end{figure}

We obtained predictions of transformed cone tip resistance throughout the domain alongside corresponding MAP estimates of the latent geological strata (Figure~\ref{fig:app_grid_z2y1}). The predicted field exhibits coherent layering, strong depth-wise gradients, and sharp transitions that coincide with major stratigraphic boundaries. The MAP geological strata exhibit laterally continuous units that closely match the overall structure of the ground model (see Figure~\ref{fig:app_grid_z1}), whilst eliminating many of the local inconsistencies and artefacts present in the ground model. Improvements are most evident in the intermediate strata GT3, GT4, and GT5, where GeoMix suppresses speckling and produces cleaner, more geologically consistent boundaries that better reflect the CPT data. GeoMix also avoids the excessive vertical smoothing often observed in standard Gaussian process models while preserving physically interpretable latent strata that are often lost in unconstrained mixture formulations.

Vertical cross-sections through the central region of the domain (Figure~\ref{fig:app_cross}; Lines~1 and~2, with Lines~3–8 in Figure~\ref{supp:fig:crossB} of the Supplementary Material) further illustrate the interaction between the inferred stratigraphy and the predicted geotechnical field. The posterior mean $q_c$ fields display pronounced layer-aligned structure, with systematic variations in cone resistance that closely follow the MAP stratigraphic boundaries. The posterior predictive standard deviation highlights elevated uncertainty at greater depths, at increasing distances from CPT profiles, and in regions where strata are thin or competing configurations are plausible. Units GT4 and GT5* exhibit greater variability throughout the domain.

\begin{figure}[t]
    \centering
    \includegraphics[]{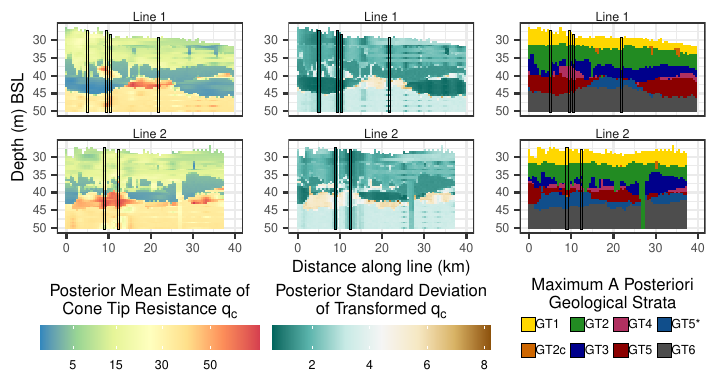}
        \caption{Predictions along two of the eight representative cross-sections through the IJV–AB site (locations shown in Figure~\ref{fig:map}) with the remaining cross sections shown in Figure~\ref{supp:fig:crossB} of the Supplementary Material. For each cross-section, the left column shows the posterior mean estimate of cone tip resistance $q_c$ and the middle column shows the predictive standard deviation, both on the transformed Box--Cox ($\lambda = 0.6$) scale, and the right column shows the maximum-a-posteriori (MAP) geological strata. Rectangles indicate observed CPT locations.}
    \label{fig:app_cross}
\end{figure}

\subsubsection{Predictive performance and model comparison}
\label{sec:app-results_test}

\begin{figure}[t]
    \centering
    \includegraphics[]{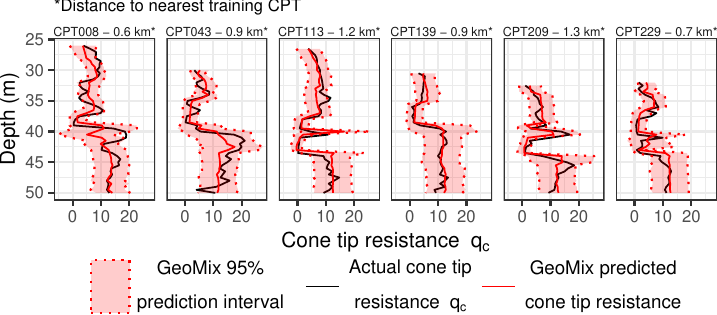}
    \caption{Six randomly selected test-set CPT profiles showing the observed cone tip resistance (black) and the GeoMix posterior mean prediction (red) with 95\% prediction intervals, all shown on the Box--Cox-transformed scale ($\lambda = 0.6$). The distance to the nearest training CPT is also marked for each example test location.}
    \label{fig:gm_test}
\end{figure}

Test set CPT profiles (Figure~\ref{fig:gm_test}) illustrate the predictive behaviour of GeoMix at six randomly selected test locations. The posterior mean (red) closely tracks the held-out transformed CPT resistance (black), and the 95\% prediction intervals provide well-calibrated uncertainty on the Box--Cox-transformed scale ($\lambda = 0.6$). Although the posterior mean is a smoothed quantity, it nevertheless reproduces the sharp depth-wise changes that characterise CPT data near stratigraphic boundaries, with tighter prediction intervals where the stratigraphic interface is confidently identified and broader intervals where boundary location is uncertain. 

Table~\ref{tab:metrics} summarises predictive performance across the competing models considered previously in the simulation study (Section~\ref{sec:sim-models}). All models exhibit negative bias, unlike the near-zero bias observed in the simulation study, likely reflecting the slightly heavier right tail of cone tip resistance values on the Box--Cox scale. Baseline GP and GeoWarp models perform relatively poorly without the ground model; incorporating $\Zvec_1$ substantially improves predictive accuracy (RMSPE) for both. Among models without $\Zvec_1$, LGFM improves on the GP and GeoWarp baselines, indicating that the latent mixture formulation captures meaningful heterogeneity in the CPT data. However, the models incorporating $\Zvec_1$ generally achieve stronger overall predictive performance. LM$+\Zvec_1$ is competitive on marginal metrics, while GeoMix and GW$+\Zvec_1$ additionally capture joint spatial dependence, as reflected by DSS2. Although GP$+\Zvec_1$ attains the lowest RMSPE (9.339), it exhibits substantial undercoverage (72.5\%) and poor probabilistic calibration, resulting in comparatively poor interval, DSS, DSS2, and logarithmic scores. GeoMix achieves comparable predictive accuracy while maintaining substantially better uncertainty quantification and overall performance.

\begin{table}[t]
\centering
\caption{Performance metrics for each fitted model based on the posterior predictive distribution $p(\Zvec_2^{\star} \mid \Zvec_1, \Zvec_2, \Xvec, \hat{\beta})$, where $\Zvec_2^{\star}$ denotes held-out geotechnical observations at test locations, including measurement error with variance $\tau^2$. }
\label{tab:metrics}
\renewcommand{\arraystretch}{0.8}
\begin{tabular}{lccccccccc}
\toprule
Model & RMSPE & Coverage & $R^2$ & Bias & IS & CRPS & DSS & DSS2 & logS \\
\midrule
GP     & 19.901 & 0.949 & 0.204 & -0.212 & 20.359 & 2.517 & 3.993 & 3.997 & 2.922 \\
GeoWarp     & 20.264 & 0.957 & 0.189 & -0.602 & 20.173 & 2.495 & 3.905 & 3.255 & 2.881 \\
LGFM  & 16.916 & 0.977 & 0.323 & -0.484 & 19.936 & 2.272 & 3.840 & 3.665 & 2.807 \\
LM$+\Zvec_1$ & 10.069 & 0.943 & 0.597 & -0.055 & 17.383 & 1.729 & 3.311 & 3.308 & 2.599 \\
GP$+\Zvec_1$ & 9.339 & 0.725 & 0.626 & -0.102 & 28.577 & 1.733 & 5.126 & 5.099 & 5.436 \\
GW$+\Zvec_1$ & 12.976 & 0.954 & 0.481 & -0.417 & 17.090 & 1.988 & 3.482 & 2.983 & 2.670 \\
GeoMix & 9.388 & 0.934 & 0.624 & -0.274 & 15.439 & 1.595 & 3.102 & 2.940 & 2.517 \\
\bottomrule
\end{tabular}
\end{table}


\section{Conclusion}
\label{sec:con}

This paper introduced GeoMix, a fully Bayesian hierarchical framework for integrating geophysical and geotechnical observations to characterise seabed conditions at large offshore sites. GeoMix couples a Potts Markov random field prior for stratigraphic structure with class-specific Gaussian processes for geotechnical properties, and incorporates a boundary-aware misclassification model that reconciles geological strata identified by the ground model with the latent seabed structure. This construction enables coherent uncertainty propagation from heterogeneous observations to engineering-relevant predictions, addressing limitations of conventional ground model workflows based on subjective allocation of stratigraphic boundaries.

Applied to the IJV–AB site, GeoMix produces spatially coherent stratigraphic estimates that reproduce large-scale stratigraphic structure while reflecting uncertainty in regions of complex layering or sparse CPT coverage. The model delivers accurate and well-calibrated predictions of cone tip resistance, outperforming baseline GP and GeoWarp approaches and demonstrating the value of incorporating ground model information. Importantly, GeoMix provides a principled quantification of uncertainty in stratigraphic boundaries across the three-dimensional domain, identifying where layer transitions are strongly supported and where alternative configurations remain plausible. More broadly, GeoMix provides a statistical basis for probabilistic ground models in offshore engineering while naturally accommodating richer geophysical inputs, alternative spatial priors, and scalable likelihood approximations.

Several avenues for future work remain. Computationally, full MCMC inference remains expensive, motivating faster alternatives such as variational inference and improved sampling strategies for the discrete field $Y_1(\cdot)$, such as block updates or tempering. Model extensions include anisotropic or longer-range spatial priors for the Potts field, and, more ambitiously, direct incorporation of raw geophysical measurements instead of the pre-processed ground model, enabling fully probabilistic joint inversion of stratigraphy and geotechnical properties.

\if1\anon
{
\section{Acknowledgments}\label{acknowledgments}
Fraser Bransby holds the Fugro Chair in Geotechnics at UWA, supported by Fugro Australia Marine. Phil Watson leads the Shell Chair in Offshore Engineering research team at UWA, supported by Shell Australia. We thank Dr Ulysse Lebrec for helpful feedback on the geophysical investigation. We also acknowledge the TIDE industry partners (INPEX Operations Australia, Shell Australia, Woodside Energy, Fugro Australia Marine, Wood Group Kenny Australia, RPS Group, Bureau Veritas, and Lloyd's Register Global Technology) for their support. Generative AI tools such as ChatGPT (5.5) and Claude Code (Sonnet 4.6) were used to assist in reviewing paper language and code development under rigorous review. 

\section{Funding}\label{funding}
This research was supported by the ARC ITRH for Transforming Energy Infrastructure through Digital Engineering (TIDE, http://TIDE.edu.au), which is led by The University of Western Australia (UWA), delivered with The University of Wollongong and a number of Australian and international research partners, and funded by the Australian Research Council, INPEX Operations Australia, Shell Australia, Woodside Energy, Fugro Australia Marine, Wood Group Kenny Australia, RPS Group, Bureau Veritas, and Lloyd's Register Global Technology (grant No. IH200100009). 
}\fi

\section{Disclosure statement}\label{disclosure-statement}
The authors declare no conflicts of interest.

\section{Data Availability Statement}
\label{data-availability-statement}
The site investigation data of Section~\ref{sec:app} cannot be shared publicly due to commercial confidentiality. However, the simulation study data of Section~\ref{sec:sim} and the code for GeoMix will be made publicly available on GitHub upon publication.

\phantomsection\label{supplementary-material}
\bigskip

\begin{center}
{\large\bf SUPPLEMENTARY MATERIAL}
\end{center}

\begin{description}
\item[Supplementary material:]
Additional simulation details, supporting figures, tables, and MCMC diagnostics. (PDF file)
\item[geomix R package] R package implementing the GeoMix model (.zip).
\item[geomix paper repository] Reproducible analysis repository containing scripts for the simulation study and application. Simulation study data are included (.zip).
\end{description}


\clearpage
\input{arxiv-supp-body}
\clearpage
\bibliography{references}
\end{document}

%% file: arxiv-supp-body.tex
\setcounter{section}{0}
\renewcommand{\thesection}{S\arabic{section}}
\setcounter{figure}{0}
\renewcommand{\thefigure}{S\arabic{figure}}
\setcounter{table}{0}
\renewcommand{\thetable}{S\arabic{table}}
\setcounter{equation}{0}
\renewcommand{\theequation}{S\arabic{equation}}
\renewcommand{\theHsection}{supp.\arabic{section}}
\renewcommand{\theHfigure}{supp.\arabic{figure}}
\renewcommand{\theHtable}{supp.\arabic{table}}
\renewcommand{\theHequation}{supp.\arabic{equation}}
\spacingset{1.9}

\spacingset{1.9} 
\captionsetup{
  font={stretch=1}
}

\begin{center}
    {\LARGE\bfseries Supplementary Material}\\[1em]
    {\Large\bfseries
    GeoMix: A Bayesian Framework for Stratigraphic\\
    and Geotechnical Inference}
\end{center}

We use the following notation in the online supplement. (1), Algorithm~1,
Section~1, etc., refer to the main paper, while (S1), Algorithm~S1, Section~S1, etc, refer to the supplement. All the acronyms used without definition in the supplement are defined in the main paper.

We organise the Supplementary Material as follows.
Section~\ref{supp:figures} presents additional tables and figures supporting the main paper.
Section~\ref{supp:boundary-terms} provides further details and contextual justification for the boundary-aware misclassification mechanism introduced in Section~\ref{sec:model-Z1}.
Section~\ref{supp:grouped-vecchia} describes the grouped Vecchia approximation used in GeoMix, including the construction of groups and conditioning sets.
Section~\ref{supp:sec:geomix-mcmc} presents the full GeoMix MCMC sampler as a step-by-step algorithm.
Section~\ref{supp:gibbs-detail} contains detailed derivations for each component of \eqref{eq:full_conditional_Y1}.
Section~\ref{supp:prediction} outlines the procedure for obtaining predictions.
Section~\ref{supp:sim-eresults} presents an additional simulation study analysis assessing the model's ability to recover the true parameter values.
Finally, Sections~\ref{supp:sim-mcmc} and \ref{supp:ijv-mcmc} present MCMC diagnostics for the simulation study and the IJV wind farm application, respectively.
\clearpage
\section{Additional Tables and Figures}
\label{supp:figures}
\noindent\mbox{}\par

 \begin{table}[H]
\centering
\caption{Summary of the principal geological strata interpreted across the IJV-AB site, which form the basis of the stratigraphic labels of the ground model used throughout the modelling.}
\label{supp:tab:geounits}
\renewcommand{\arraystretch}{1.1}
\begin{tabularx}{\linewidth}{P{1.1cm} P{3.0cm} Y}
\toprule
\textbf{Unit} & \textbf{Geological Name} & \textbf{Description} \\
\midrule
GT1  & Bligh Bank Formation & Fine-medium sand forming the upper seabed, and forming a series of tidal ridges and sand waves. Widespread and 0.3-8.9\,m thick. \\[3pt]

GT2c & Holocene (Undifferentiated) Channels & Infilled channels of fine-medium sand, sandy clay, and silt with organics. Found mostly in the north, averages around 5m thick although can vary between 0.5-24.5\,m thick. \\[3pt]

GT2  & Naaldwijk Formation & Well-sorted fine-medium sand, medium-dense to dense, thickening in SE-NW channels. Laterally extensive, averaging 4.75\,m thick and varying between 0.5-17.5\,m thick. \\[3pt]

GT3  & Eem (Brown Bank) Member & Silty to sandy clay with thin sand lenses. Cohesive, compressible, and discontinuous, typically 0.5-8\,m thick. \\[3pt]

GT4  & Eem Sand Member & Fine-medium dense sand forming a thin, laterally discontinuous layer within the Eem sequence, up to 4.5\,m thick. \\[3pt]

GT5  & Eem Clay Member & Sandy to gravelly clay/silt of low permeability and moderate plasticity. Cohesive confining layer 0.5-12\,m thick. \\[3pt]

GT5* & Transitional Soil Member & Dense fine-medium sand grading between cohesive and granular units. Channelised, discontinuous, 0.5-6\,m thick. \\[3pt]

GT6  & Yarmouth Roads Formation & Dense interbedded sand, silt, and clay with occasional peat. Regionally extensive from near seabed to beyond 80\,m depth. \\
\bottomrule
\end{tabularx}
\end{table}

\begin{figure}[H]
    \centering
    \includegraphics{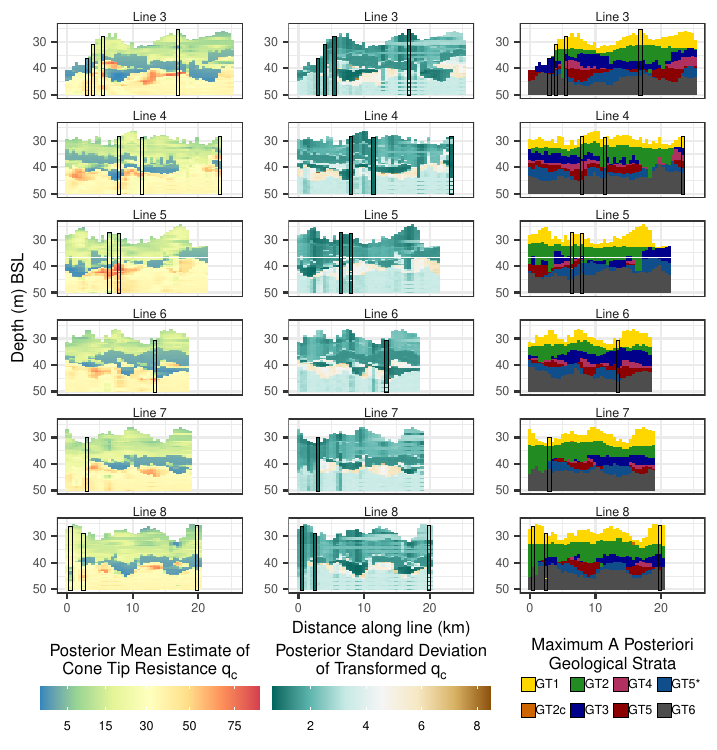}
        \caption{Predictions along Lines 3-8 (see Figure \ref{fig:map}) demonstrating additional representative cross-sections through the IJV–AB site. For each cross-section, the left column shows the posterior mean estimate of cone tip resistance $q_c$, the middle column shows the prediction standard deviation on the transformed Box-Cox ($\lambda = 0.6$) scale, and the right column shows the maximum-a-posteriori (MAP) geological strata. Rectangles indicate observed CPT locations intersecting each cross-section.}
    \label{supp:fig:crossB}
\end{figure}

\section{Boundary-aware misclassification mechanism}
\label{supp:boundary-terms}
This section provides the mathematical formulation of the boundary-aware misclassification mechanism introduced in Section~\ref{sec:model-Z1}, including the construction of boundary influence terms and the specification of confusion probabilities that govern the likelihood of observed ground model strata $\Zvec_1$.
\subsection{The boundary influence term}
Consider a set of locations $\Omega \subseteq \mathcal{S}_1$, that contains points along a depth transect at a single horizontal location and denote these locations as $\svec_{1,1},\dots,\svec_{1,w}$, for $w \ge 2$.

Define the class-membership indicator $I_{i,\ell}=1$ if $Y_1(\svec_{1,i})=\ell$ and $0$ otherwise, for $i=1,\ldots,w$, $\ell=1,\ldots,K$.
We encode adjacent class boundaries for class $\ell$ by
\[
J_{r,\ell}=|I_{r,\ell}-I_{r+1,\ell}|,\qquad r=1,\ldots,w-1.
\]
Let $d_i$ denote the corresponding depth coordinate at location $\svec_{1,i}\in\Omega$.
For a bandwidth parameter $h>0$, we construct a local depth kernel
\[
\phi_{r, i}
\;=\;
\Phi\!\left(
-\frac{1}{h}\left| \frac{d_{r+1} + d_{r}}{2} - d_{i}   \right|
\right),
\qquad r = 1,\ldots,w-1,\ \ i = 1,\ldots,w,
\]
where $\Phi(\cdot)$ denotes the standard normal cumulative distribution function.
The resulting local depth kernel matrix $\boldsymbol\Phi = (\phi_{r,i} : r = 1,\ldots,(w-1); i = 1,...,w)$ has dimension $(w-1)\times w$, with rows corresponding to midpoints between adjacent depth grid points $d_{r},d_{r+1}$, and columns corresponding to depth indices $d_{i}$. Midpoints are used to represent potential interfaces between class labels, since the latent class is assumed constant within each voxel centred at $d_i$, and transitions between classes can occur only along the voxel boundaries, which correspond to these midpoints.

We use this kernel to construct a boundary-aware representation that adjusts the probability of misclassifying $Z_{1,i}$ relative to its true class $Y_{1,i}$, for $i=1,\ldots,N_1$.
This formulation ensures that misclassification uncertainty concentrates at interfaces rather than within homogeneous regions.
Now, recall that $J_{r,\ell} = 1$ indicates the presence of a class-$\ell$ boundary between adjacent depths $d_{r}, d_{r+1}$; we smooth out this indicator with the kernel matrix $\boldsymbol\Phi$ to generate a basis expansion for what we term ``misclassification influences'' as follows,
\begin{equation}
\widetilde{\varphi}_{i,\ell}=\sum_{r=1}^{w-1} \phi_{r,i}J_{r,\ell},\qquad
\varphi_{i,\ell}=\min\{\widetilde{\varphi}_{i, \ell},\,1\}\in[0,1],
\label{supp:eq:varphi}
\end{equation}
for $i = 1,...,w$ and $\ell = 1,...,K$.

Here, $\varphi_{i,\ell}$ quantifies the influence of a class-$\ell$ boundary on misclassification at depth $d_i$, where proximity is defined through the bandwidth parameter $h\geq0$.
A large value of $\varphi_{i,\ell}$ corresponds to $d_i$ being ``near'' one or more boundaries involving class $\ell$.
Contextually, we expect a given geological stratum (class $\ell$) to exhibit at most two boundaries: one where the depth profile enters the class and one where it exits.
Accordingly, $\varphi_{i,\ell}\approx 0.5$ when $d_i$ is close to a single boundary, and $\varphi_{i,\ell}\approx 1$ when $d_i$ is close to both boundaries; this occurs when strata are only one voxel thick (i.e., when $Y_1(\svec_{1,i-1})\neq Y_1(\svec_{1,i})\neq Y_1(\svec_{1,i+1})$).
This construction reflects the intuition that there is a greater chance of misclassifying $Y_1(\cdot)$ when strata are not thick.

To ensure the boundary influence is bounded in all cases, we cap $\varphi_{i,\ell}$ at one.

\subsection{Confusion probabilities}
Let $\gamma_{k,\ell}$ denote weights, with $k = 1,\ldots,K$ denoting the latent class and $\ell = 1,\ldots,K$ denoting the observed class, where $\gamma_{k,\ell} \in (0,1)$ for $k\neq\ell$ and $\gamma_{k,k}=0$. Let $\kappa \in [0,1]$ denote a baseline intercept controlling the probability of correct classification. We introduce $\kappa$ to force a minimum classification probability across all classes, namely $\delta_\kappa = (1-\kappa)/(K-1)$.

We require $\sum_{\ell=1}^K \gamma_{k,\ell} \leq 1$ for $k=1,\ldots,K$.
In addition, recall that the boundary influence terms $\varphi_{i, \ell}\in[0,1]$ are capped at one.
Together, these restrictions ensure that
$
\sum_{\ell\neq k}\gamma_{k,\ell}\,\varphi_{i, \ell}\leq 1,
$
so that the following probabilities are well defined.

For location $\svec_{1,i}$ at a given depth transect region $\Omega$ of size $w$, with true class $Y_1(\svec_{1,i}) = k$, we define the class probability vector $\boldsymbol{\pi}_{i,k}\equiv(\pi_{i,k,1},\ldots,\pi_{i,k,K})'$ as
\begin{equation}
\pi_{i,k,\ell}=
\begin{cases}
\displaystyle \kappa-\kappa\sum_{\ell'\neq k}\gamma_{k,\ell'}\,\varphi_{i, \ell'}, & \ell=k,\\[10pt]
\displaystyle \frac{1-\kappa}{K-1}+\kappa\,\gamma_{k,\ell}\,\varphi_{i, \ell}, & \ell\neq k,
\end{cases}
\qquad i=1,\ldots,w,\ \ \ell=1,\ldots,K.
\label{supp:eq:pi}
\end{equation}

This specification captures a key feature of the application: away from the boundaries (where $\varphi_{i, \ell}$ is small), the probability of misclassification converges to the baseline level $\frac{1-\kappa}{K-1}$; near the boundaries it increases, with the pattern determined by neighbouring classes $\ell$ through their basis function weights $\gamma_{k,\ell}$.
The expression $\frac{1-\kappa}{K-1}$ thus provides a lower bound on misclassification and $\kappa$ bounds the correct classification probability from above, ensuring that even in homogeneous regions there remains a nonzero probability of observation error.

\section{Grouped Vecchia factorisation}
\label{supp:grouped-vecchia}
To render computation tractable over large spatial domains,
we employ a grouped Vecchia approximation,
which expresses the joint distribution of the observations into a product of conditional densities evaluated group by group \citep{vecchia1988,Katzfuss2021}.
We refer to this as a \emph{grouped} Vecchia approximation rather than a \emph{block} Vecchia scheme to emphasise that the grouping structure need not coincide with the spatial lattice or a specific ordering of locations.

Recall from Section \ref{sec:model-Y2} that $\mathcal{S}_2 \subseteq \mathcal{S}_1$ denotes the set of spatial locations at which continuous measurements $\Zvec_2$ are observed, and that $\mathcal{S}_2^{(k)}$ denotes the subset of locations belonging to class $k$, for $k=1,\ldots,K$.

Suppose that each location $\svec_2 \in \mathcal{S}_2$ is assigned to one of $M$ non-overlapping groups $\Hs_g$, indexed by $g=1,\ldots,M$, such that
\[
\mathcal{S}_2 = \bigcup_{g=1}^{M} \Hs_g,
\qquad
\{\Hs_g\}_{g=1}^{M} \text{ partitions } \mathcal{S}_2.
\]

For each class $k=1,\ldots,K$ and group $g=1,\ldots,M$, define the class--group intersection
\[
\Gs_{k,g} \equiv
\left\{\, \svec_2 \in \Hs_g :
Y_1(\svec_2)=k \right\}.
\]

By construction,
\[
\mathcal{S}_2^{(k)} = \bigcup_{g=1}^{M} \Gs_{k,g},
\quad
\{\Gs_{k,g}\}_{g=1}^{M} \text{ partitions } \mathcal{S}_2^{(k)},
\quad
\Hs_g = \bigcup_{k=1}^{K} \Gs_{k,g}.
\]

Thus the collection
$\{\Gs_{k,g}\}_{k=1,\ldots,K;\,g=1,\ldots,M}$
simultaneously defines both the class-specific partitions $\mathcal{S}_2^{(k)}$ and the group partition $\{\Hs_g\}_{g=1}^{M}$ of $\mathcal{S}_2$.

For each group $g=1,\ldots,M$, define a class-agnostic conditioning set
\[
C(\Hs_g) \subseteq \bigcup_{j=1}^{g-1} \Hs_j,
\qquad |C(\Hs_g)| \le m,
\]
where $m$ denotes the Vecchia conditioning size.

Class-specific conditioning sets are obtained by restriction,
\[
C(\Gs_{k,g})
\equiv
\left\{\, \svec_2 \in C(\Hs_g) :
Y_1(\svec_2)=k \right\}
\subseteq C(\Hs_g),
\qquad |C(\Gs_{k,g})| \le m.
\]

For any subset $A \subseteq \mathcal{S}_2$, let
\[
\Zvec_{2}^{(A)} \equiv \left( Z_{2,i} : \svec_{2,i} \in A \right)'
\]
denote the corresponding subvector of $\Zvec_2$.

Under the grouped Vecchia construction,
\[
p\!\left(\Zvec_2 \mid \Yvec_1, \Xvec, \thetab_{2}\right)
\approx
\prod_{g=1}^{M}
p\!\left(
\Zvec_2^{(\Hs_g)}
\,\middle|\,
\Zvec_2^{(C(\Hs_g))},\,\Yvec_1,\,
\Xvec,\,
\thetab_{2}
\right).
\]

Because the Gaussian process components are conditionally independent across classes given $\Yvec_1$, each group contribution further factorises over $k$, yielding
\begin{equation}
p\!\left(\Zvec_2 \mid \Yvec_1, \Xvec, \thetab_{2}\right)
\approx
\prod_{g=1}^{M}
\ \prod_{k=1}^{K}
p\!\left(
\Zvec_2^{(\Gs_{k,g})}
\,\middle|\,
\Zvec_2^{(C(\Gs_{k,g}))},\, \Yvec_1, \,
\Xvec,\,
\thetab_{2}
\right).
\label{supp:eq:Z2-likelihood}
\end{equation}

We now provide a closed-form expression for the approximate likelihood of the continuous measurements $\Zvec_2$ induced by the grouped Vecchia factorisation.
Each conditional factor follows from the class-specific Gaussian-process specification, yielding a tractable approximation to the marginal likelihood $p(\Zvec_2 \mid \Yvec_1, \Xvec, \thetab_2)$.

For $A,B \subseteq \mathcal{S}_2^{(k)}$, define the mean and covariance blocks
\[
\mub^{(k)}_A \equiv \{\xvec_a' \alphab_k\}_{a\in A},
\qquad
\Sigmab^{(k)}_{A,B} \equiv \operatorname{Cov}\!\big(\Zvec_2^{(A)},\Zvec_2^{(B)}\big).
\]

Let $\Gs_{k,g}$ denote the $g$th Vecchia group for class $k$ and $\Cs_{k,g}\equiv C(\Gs_{k,g})$ its conditioning set.
Each factor in the Vecchia approximation is multivariate normal,
\[
\Zvec_2^{(\Gs_{k,g})} \,\big|\, \Zvec_2^{(\Cs_{k,g})}, \Xvec,\thetab_{2}
\;\sim\; \mathcal{N}\!\bigl(\mathbf{m}_{k,g},\, \Sigmab_{k,g}\bigr),
\]
with mean
\[
\mathbf{m}_{k,g}
= \mub^{(k)}_{\Gs_{k,g}}
+ \Sigmab^{(k)}_{\Gs_{k,g},\Cs_{k,g}}
\bigl(\Sigmab^{(k)}_{\Cs_{k,g},\Cs_{k,g}}\bigr)^{-1}
\!\bigl(\Zvec_2^{(\Cs_{k,g})} - \mub^{(k)}_{\Cs_{k,g}}\bigr),
\]
and covariance
\[
\Sigmab_{k,g}
= \Sigmab^{(k)}_{\Gs_{k,g},\Gs_{k,g}}
- \Sigmab^{(k)}_{\Gs_{k,g},\Cs_{k,g}}
\bigl(\Sigmab^{(k)}_{\Cs_{k,g},\Cs_{k,g}}\bigr)^{-1}
\Sigmab^{(k)}_{\Cs_{k,g},\Gs_{k,g}} .
\]
When $\Cs_{k,g}=\emptyset$, this reduces to
$\mathbf{m}_{k,g}=\mub^{(k)}_{\Gs_{k,g}}$ and
$\Sigmab_{k,g}=\Sigmab^{(k)}_{\Gs_{k,g},\Gs_{k,g}}$.

Hence the approximate log-likelihood for $Z_2$ under grouped Vecchia is
\begin{equation}
\log p\!\left(\Zvec_2 \mid \Yvec_1, \Xvec, \thetab_2\right)
= -\tfrac12 \sum_{k=1}^K \sum_{g=1}^{M}
\Bigl\{ \log\!\det(\Sigmab_{k,g})
+ {\tilde{\Zvec}_2^{(\Gs_{k,g})\prime}} \Sigmab_{k,g}^{-1} \tilde{\Zvec}_2^{(\Gs_{k,g})}
+ |\Gs_{k,g}| \log(2\pi) \Bigr\},
\label{supp:eq:Z2-vecchia}
\end{equation}
where $\tilde{\Zvec}_2^{(\Gs_{k,g})}\equiv \Zvec_2^{(\Gs_{k,g})} - \mathbf{m}_{k,g}$.
In the computation, we evaluate $\log\!\det(\Sigmab_{k,g})=2\sum_i \log (\mathbf{L}_{k,g})_{ii}$ and ${\tilde{\Zvec}_2^{(\Gs_{k,g})\prime}} \Sigmab_{k,g}^{-1} \tilde{\Zvec}_2^{(\Gs_{k,g})} = \|\mathbf{L}_{k,g}^{-1}\tilde{\Zvec}_2^{(\Gs_{k,g})}\|_2^2$ using the Cholesky factor $\Sigmab_{k,g}=\mathbf{L}_{k,g}\mathbf{L}_{k,g}'$.

\subsection{Group construction and conditioning sets in the Vecchia approximation}

In GeoMix, the grouped Vecchia approximation is constructed using depth-based groups so that the dependence structure aligns with the vertical organisation of the sedimentary profile. Observations are partitioned into groups according to discrete depth bins, with all observations at the same depth assigned to the same group. The groups are ordered from the shallowest depth (closest to the seabed surface) to the deepest. This ordering reflects the strong vertical structure in the data and provides a natural sequential structure for the approximation.

Let $g=1,\ldots,M$ index these ordered depth groups. As mentioned previously, the Vecchia approximation factorises the joint distribution of the observations into a product of conditional densities evaluated group by group. Because the group sizes are larger than the conditioning size $m$, each group conditions only on observations from the immediately preceding depth group. Consequently, observations in group $g$ depend only on a subset of observations from group $g-1$, while the first group is unconditional.

For each group $g>1$, the conditioning set is constructed from observations in the preceding group. Recall $\mathcal{H}_g$ denotes the observations in group $g$ irrespective of class, and hence $\mathcal{H}_{g-1}$ those in the preceding group. Distances between locations in $\mathcal{H}_g$ and $\mathcal{H}_{g-1}$ are computed using rank-transformed spatial distances. For each observation in $\mathcal{H}_g$, candidate neighbours are defined as its $q$ nearest observations in $\mathcal{H}_{g-1}$, where $q \ge 2$ is chosen as the smallest value such that the union of these neighbours across the group contains at least $m$ unique candidates. This ensures that all observations in the group contribute potential conditioning points while avoiding an excessively large candidate set.

If the resulting candidate set contains exactly $m$ observations, all candidates are retained. Otherwise, the candidate set is reduced to size $m$ using a coverage-based selection procedure. Specifically, let $D_{c,u}$ denote the distance between candidate location $c \in \mathcal{H}_{g-1}$ and observation $u \in \mathcal{H}_g$. We select a subset $S$ of $m$ candidates to approximately minimise the maximum distance from any observation in the current group to its nearest selected candidate,
\[
\min_{|S|=m} \max_{u \in \mathcal{H}_g} \min_{c \in S} D_{c,u}.
\]
This is implemented using a greedy $k$-centre selection algorithm \citep{gonzalez1985}. The resulting conditioning set therefore provides good spatial coverage of the current group while maintaining the fixed size $m$.

This construction ensures that each observation depends only on observations within its own depth group and a small, representative subset of the layer immediately above it. Such nearest-neighbour conditioning schemes are widely used in scalable Gaussian process approximations \citep{Katzfuss2021}. As a consequence, when the latent geological class at a location changes, only the likelihood contributions of the affected group and its neighbouring group need to be recomputed, substantially reducing the computational burden of the MCMC algorithm.

\section{GeoMix MCMC Sampler}
\label{supp:sec:geomix-mcmc}

\begin{algorithm}[H]
\small
\caption{GeoMix MCMC sampler.}
\label{supp:alg:geomix-mcmc}
\begin{algorithmic}[1]
\Require Observed data $(\Zvec_1,\Zvec_2,\Xvec)$, lattice $\mathcal{S}_1$ with subset $\mathcal{S}_2$, grouped Vecchia groups and conditioning sets, fixed Potts interaction $\hat{\beta}$, priors for $(\thetab_1,\thetab_2)$, and iteration size $B$.
\State Initialise $\Yvec_1^{(0)}$, $\thetab_1^{(0)}$, and $\thetab_2^{(0)}$.

\For{$b = 1,\ldots,B$}

    \Statex \emph{Update latent class field $\Yvec_1$ (Gibbs)}
    \For{each location $\svec_{1,i} \in \mathcal{S}_1$}
        \State Compute $\Pr(Y_{1,i}=k \mid \cdot)$ for $k=1,\ldots,K$ using \eqref{eq:full_conditional_Y1}.
        \State Sample $Y_{1,i}^{(b)}$ from the resulting categorical distribution.
    \EndFor

    \Statex \emph{Update regression coefficients $\alphab_k$ (Gibbs)}
    \For{$k = 1,\ldots,K$}
        \State Sample $\alphab_k^{(b)}$ from its full conditional distribution.
    \EndFor

    \Statex \emph{Update Gaussian process parameters (HMC)}
    \State Update $\sigma_L^{2\,(b)}$ and $\sigma_D^{2\,(b)}$.
    \For{$k = 1,\ldots,K$}
        \State Update $\sigma_k^{2\,(b)}$, $\ell_{L,k}^{(b)}$, and $\ell_{D,k}^{(b)}$.
    \EndFor

    \Statex \emph{Update basis function weights $\gamma_{k,\ell}$, for $k,\ell = 1,\ldots,K$ (Metropolis--Hastings) and confusion bandwidth parameter $h$}
    \For{$k = 1,\ldots,K$}
        \State Propose $\boldsymbol{\eta}_k^{\ast}$ from a Dirichlet random-walk proposal.
        \State Compute corresponding $\gamma_{k,\ell}^{\ast}$ for $\ell=1,\ldots,K$ with $\gamma_{k,k}^{\ast}=0$.
        \State Accept or reject using the Metropolis-Hastings acceptance probability.
    \EndFor
    \State Update $h^{(b)}$ via HMC.

    \Statex \emph{Posterior prediction for $Y_2(\cdot)$ (optional)}
    \State Draw from $p\!\left( Y_2(\cdot)\mid \Zvec_2, \Yvec_1^{(b)}, \thetab_2^{(b)}, \Xvec \right)$ using the grouped Vecchia approximation.

\EndFor

\State \Return Posterior samples $\{\Yvec_1^{(b)}, \thetab_1^{(b)}, \thetab_2^{(b)}\}_{b=1}^B$ and posterior predictive draws for $Y_2(\cdot)$.
\end{algorithmic}
\end{algorithm}

\section{Gibbs Updates of the Latent Geological Strata}
\label{supp:gibbs-detail}
In this section we outline the detailed derivation required for each component of \eqref{eq:full_conditional_Y1}.

\subsection{Potts prior: Single-location update}
\label{supp:gibbs-Y1-potts}
Spatial coherence in the latent field $\Yvec_1$ is encoded via a Potts Markov random field over the lattice $\Ss_1$. By the local Markov property, the full conditional for each site depends only on the class labels at adjacent locations, as given in \eqref{eq:potts}.

\subsection{Boundary-aware misclassification likelihood: Single-location update}
\label{supp:gibbs-Y1-boundary}
Recall from~\eqref{eq:Z1-likelihood} that the likelihood of $\Zvec_1$ is constructed under the assumption of location-wise conditional independence given $\Yvec_1$ and $\thetab_1$. The likelihoods of distinct vertical profiles are conditionally independent by design, since the boundary-aware misclassification mechanism is defined within, but not across, vertical profiles.

Therefore, for a proposed update $Y_{1,i}=k$ at location $\svec_{1,i}\in\mathcal{S}_1$ belonging to a vertical profile $\mathcal{V}$, the conditional likelihood with respect to $Y_{1,i}=k$, holding all other values fixed, is proportional to the likelihood of the observations within that profile, $\Zvec_1^{(\mathcal{V})} \equiv (Z_{1,i} : \svec_{1,i}\in\mathcal{V})'$, that is,
\[
p\big(\Zvec_1 \mid Y_{1,i}=k,\, \Yvec_{1}^{\setminus i},\, \thetab_1\big)
\propto
p\big(\Zvec_1^{(\mathcal{V})} \mid Y_{1,i}=k,\, \Yvec_{1}^{\setminus i},\, \thetab_1\big),
\qquad \text{w.r.t.\ } k.
\]
However, since all class labels in $\mathcal{V}$ remain fixed except at the update location $\svec_{1,i}$, recomputing the entire profile likelihood at every iteration is computationally burdensome. To address this, we use a \emph{localised approximation} in which the likelihood is evaluated only for a neighbourhood (or window) of nearby depths instead of the full profile.

\subsubsection{Local approximation using a window for single-location updates}
Let $\svec_{1,i} \equiv (x, y, d_i)$ denote the location being updated, and let the vertical profile at lateral location $(x,y)$ be
$\mathcal{V} \equiv \{\svec_{1,j_0}, \ldots, \svec_{1,j_0+|\mathcal{V}|-1}\}$ for some known starting index $j_0$.
We define a centred depth-wise window of size at most $2a+1$, where $a \in \mathbb{N}$ is a fixed half-width with $a < |\mathcal{V}|$.
Let
\[
\mathcal{J}_i \equiv \big\{ \max(i-a,\,j_0),\ \ldots,\ i,\ \ldots,\ \min(i+a,\, j_0+|\mathcal{V}|-1) \big\},
\]
and define the associated local window of locations
\[
\Omega_i \equiv \{\, (x,y,d_j) \in \mathcal{S}_1 : j \in \mathcal{J}_i \,\}.
\]
Within this window $\Omega_i$, the misclassification likelihood is approximated by restricting computation to the class labels and observations indexed by $\mathcal{J}_i$.

To justify this approximation, recall that each boundary term $\varphi_{i,\ell}$ is constructed by smoothing discrete boundary indicators $J_{r,\ell}$ using the kernel
\[
\phi_{r,i}
= \Phi\!\left(-\frac{1}{h}\,\Big|\,\tfrac{d_{r}+d_{r+1}}{2} - d_{i}\Big|\right), \qquad r,i \in \mathcal{V}.
\]
On a semi-regular depth grid with spacing $\Delta_d \approx d_{i+1}-d_i$, we observe
\[
\Big|\tfrac{d_{r}+d_{r+1}}{2} - d_{i}\Big|
\approx b\,\Delta_d, \qquad
b \equiv \big|r - i + \tfrac{1}{2}\big| \in \mathbb{Z}+\tfrac{1}{2},
\]
so that
\[
\phi_{r,i} = \Phi\!\left(-\frac{b\,\Delta_d}{h}\right),
\]
which decays exponentially as $b \to \infty$.

Let $\pi^{*}_{i,k,\ell}$ denote the confusion probability obtained by replacing $\varphi_{i,\ell}$ with its window-based approximation $\varphi^{*}_{i,\ell}$.
From Section~\ref{sec:model-Z1},
\[
\pi_{i,k,\ell} =
\begin{cases}
\tfrac{1-\kappa}{K-1} + \kappa\,\gamma_{k,\ell}\,\varphi_{i,\ell}, & k\neq \ell, \\[4pt]
\kappa - \kappa\sum_{\ell'\neq k}\gamma_{k,\ell'}\,\varphi_{i,\ell'}, & k=\ell,
\end{cases}
\]
and $\pi^{*}_{i,k,\ell}$ is defined analogously with $\varphi_{i,\ell}$ replaced by $\varphi^{*}_{i,\ell}$.

Because $\phi_{r,i}$ decays in $|r-i|$, the influence of a label change at $\svec_{1,r}$ on boundary weights $\varphi_{i,\ell}$ is negligible ($<\delta$) for all $i \notin \mathcal{J}_i$ provided
\[
\phi_{r,i} \le \Phi\!\left(-\frac{(a+1)\Delta_d}{h}\right).
\]
A sufficient condition is therefore
\[
a \ge \left\lceil \frac{h}{\Delta_d}\,\big|\Phi^{-1}(\delta)\big| \right\rceil - 1,
\]
which ensures $|\varphi_{i,\ell} - \varphi^{*}_{i,\ell}| \le \delta$  for all $\ell$.

It follows that under this choice of $a$,
for $k\neq \ell$,
\[
|\pi_{i,k,\ell}-\pi^{*}_{i,k,\ell}|
= \kappa\,\gamma_{k,\ell}\,|\varphi_{i,\ell}-\varphi^{*}_{i,\ell}|
< \delta,
\]
since $\kappa,\gamma_{k,\ell} \le 1$;
and for $k=\ell$,
\[
|\pi_{i,k,k}-\pi^{*}_{i,k,k}|
= \kappa\sum_{\ell'\neq k}\gamma_{k,\ell'}\,|\varphi_{i,\ell'}-\varphi^{*}_{i,\ell'}|
< \kappa\sum_{\ell'\neq k}\gamma_{k,\ell'}\,\delta
\le \delta,
\]
because $\kappa < 1$ and $\sum_{\ell'\neq k}\gamma_{k,\ell'} \le 1$. Note the constraint $\sum_{\ell'\neq k}\gamma_{k,\ell'} \leq 1$ holds uniformly across all MCMC iterations as under the $\mathrm{Dirichlet}(\mathbf{1}_K)$ prior, $\gamma_{k,\ell} = \eta_{k,\ell}$ for $\ell\neq k$ and $\gamma_{k,k}=0$, so $\sum_{\ell'\neq k}\gamma_{k,\ell'} = 1 - \eta_{k,k} \leq 1$ by construction.

Thus, for all $k,\ell$, $
|\pi_{i,k,\ell}-\pi^{*}_{i,k,\ell}| < \delta,$
showing that restricting computation to the local window $\Omega_i$ introduces at most $\delta$ error in each confusion probability.

This establishes that, for a given numerical tolerance $\delta > 0$, the contribution of any location outside the local window $\Omega_i$ of maximum half-width $a$, is uniformly bounded by $\delta$ and decays exponentially in squared distance.
Hence, restricting the misclassification likelihood to the local window $\Omega_i$ provides a theoretically justified and computationally efficient approximation.

\subsection{Gaussian process likelihood: Single-location update}
\label{supp:gibbs-Y1-gp}

Recall from Section~\ref{sec:model-Z2} that the continuous-valued geotechnical measurements $\Zvec_2$ are modelled using class-specific Gaussian processes, with parameters denoted $\thetab_2$.
Consider a single-location update to the latent class label at location $\svec_{1,i} \in \mathcal{S}_1$.

If $\svec_{1,i} \notin \mathcal{S}_2$, then no continuous measurements are observed at this location, and a proposed change $Y_1(\svec_{1,i}) =k$ has no effect on the likelihood of $\Zvec_2$.
In contrast, if $\svec_{1,i} \in \mathcal{S}_2$, then the likelihood of $\Zvec_2$ depends on the local class label and must be updated accordingly.
To make this case explicit, we write $\svec_{2} \in \mathcal{S}_2 \subseteq \mathcal{S}_1$ to denote a location where continuous measurements are observed, and more specifically $\svec_{2,i}$ for the location being updated.

Therefore, for a proposed update $Y_{1,i}=k$ at location $\svec_{2,i}\in\mathcal{S}_2$, the conditional likelihood with respect to $Y_{1,i}=k$, holding all other values fixed, is proportional to the likelihood of $\Zvec_2^{(k)}$ conditional on $\Xvec^{(k)}$, $\Yvec_1^{\setminus i}$, and $\thetab_2$. When using the grouped Vecchia approximation, this further reduces to only those $k$-class groups that either contain $Z_{2,i}$ or include it in their conditioning sets. That is, writing $\Zvec_2^{(G)} \equiv (Z_{2,i}:\svec_{2,i} \in G)'$ for some set $G$ of locations within the same group,
\[
p\big(\Zvec_2 \mid Y_{1,i}=k,\, \Yvec_{1}^{\setminus i},\,\Xvec,\ \thetab_2\big)
\propto \prod_{G \in \mathcal{C}(\svec_{2,i};k) }
p\big(\Zvec_2^{(G)} \mid Y_{1,i}=k,\, \Yvec_{1}^{\setminus i}, \, \Xvec,\, \thetab_2\big),
\qquad \text{w.r.t.\ } k,
\]
where $\mathcal{C}(\svec_{2,i};k) \equiv \{\Gs_{k,g} : \svec_{2,i} \in \Gs_{k,g} \text{ or } \svec_{2,i} \in C(\Gs_{k,g})\}$, with  $\Gs_{k,g} \equiv \{\, \svec_2 \in \mathcal{S}_2 : Y_1(\svec_2)=k,\ \text{group}(\svec_2)=g \,\}$, and $C(\Gs_{k,g})$ denoting the conditioning set of $\Gs_{k,g}$ (see Section~\ref{supp:grouped-vecchia} for more details).

To compute the likelihood contribution efficiently, it is useful to consider explicitly how the grouped Vecchia likelihood changes under a single relabelling. Let $k_0 \equiv Y_1(\svec_{2,i})$ denote the current class label at the location being updated. The contribution of $\Zvec_2$ to the full conditional of $Y_1(\cdot)$ is given by the change in the grouped Vecchia log-likelihood resulting from the proposed relabelling $Y_1(\svec_{2,i}) : k_0 \rightarrow k$:
\[
R_{\Zvec_2}(\svec_{2,i}; k_0, k) \equiv
\log p(\Zvec_2 \mid Y_{1,i}=k,\, \Yvec_{1}^{\setminus i},\,  \Xvec,\, \thetab_2)
- \log p(\Zvec_2 \mid \Yvec_1,\, \Xvec,\, \thetab_2).
\]

This term captures the local effect of class label changes on the GP likelihood and is evaluated efficiently using the grouped Vecchia approximation.

To compute this term, observe that location $\svec_{2,i}$ belongs to a single group index $g_i \in \{1, \ldots, M\}$, shared across classes by construction.
Under the original label $k_0$, the location is a member of group $\Gs_{k_0, g_i}$; under the proposed label $k$, it becomes a member of $\Gs_{k, g_i}$.
Let
\[
\mathcal{A}(\svec_2; k) \equiv \left\{ g : \svec_2 \in C(\Gs_{k,g}) \right\}
\]
denote the set of group indices in class $k$ whose conditioning sets include location $\svec_2$.

For any group $\Gs_{k,g}$, the corresponding log-likelihood contribution takes the form
\[
\ell_{k,g} \equiv -\tfrac{1}{2} \left\{
\log\det(\Sigmab_{k,g})
+ \tilde{\Zvec}_2^{(\Gs_{k,g})\prime}\Sigmab_{k,g}^{-1} \tilde{\Zvec}_2^{(\Gs_{k,g})}
+ |\Gs_{k,g}| \log(2\pi)
\right\},
\;\tilde{\Zvec}_2^{(\Gs_{k,g})}\equiv \Zvec_2^{(\Gs_{k,g})} - \mathbf{m}_{k,g},
\]
with $(\mathbf{m}_{k,g}, \Sigmab_{k,g})$ defined via the grouped Vecchia approximation in~\eqref{supp:eq:Z2-vecchia}.

To express likelihood changes due to the update at location $\svec_{2,i}$, we use the notation $\ell_{k,g}^{+i}$ and $\ell_{k,g}^{-i}$ to indicate that location $\svec_{2,i}$ is included in or excluded from the evaluation of group $\Gs_{k,g}$, respectively.
The location $\svec_{2,i}$ contributes directly to $\Gs_{k,g}$ when $g = g_i$, and contributes via the conditioning set when $g \in \mathcal{A}(\svec_{2,i}; k)$.
Inclusion and exclusion are defined accordingly in both contexts.

Let
\[
\mathcal{B}(\svec_{2,i}; k_0, k) \equiv \{g_i\} \cup \mathcal{A}(\svec_{2,i}; k_0) \cup \mathcal{A}(\svec_{2,i}; k),
\]
and define the unaffected contribution
\[
U \equiv \sum_{g \notin \mathcal{B}(\svec_{2,i}; k_0, k)} \;\sum_{k'=1}^K \ell_{k',g},
\]
which is invariant to the proposed update and cancels in the log-ratio.

Then the total log-likelihood of $\Zvec_2$ under the current configuration $\Yvec_1$ is
\[
\log p(\Zvec_2 \mid \Yvec_1, \Xvec, \thetab_2) =
\ell_{k_0, g_i}^{+i} + \ell_{k, g_i}^{-i}
+ \sum_{g \in \mathcal{A}(\svec_{2,i}; k_0)} \left( \ell_{k_0, g}^{+i} + \ell_{k, g}^{-i} \right) + U,
\]
while under the proposed configuration it becomes
\[
\log p(\Zvec_2 \mid \Yvec_{1}\!\big|_{Y_{1,i}=k},\, \Xvec,\, \thetab_2) =
\ell_{k_0, g_i}^{-i} + \ell_{k, g_i}^{+i}
+ \sum_{g \in \mathcal{A}(\svec_{2,i}; k_0)} \ell_{k_0, g}^{-i}
+ \sum_{g \in \mathcal{A}(\svec_{2,i}; k)} \ell_{k, g}^{+i} + U.
\]

The log-likelihood ratio resulting from the single-location update $Y_1(\svec_{2,i}): k_0 \rightarrow k$ is therefore
\begin{align}
R_{\Zvec_2}(\svec_{2,i}; k_0, k)
&= \big(\ell_{k, g_i}^{+i} - \ell_{k, g_i}^{-i}\big)
   + \big(\ell_{k_0, g_i}^{-i} - \ell_{k_0, g_i}^{+i}\big) \nonumber \\
&\quad + \sum_{g \in \mathcal{A}(\svec_{2,i}; k)} \big(\ell_{k, g}^{+i} - \ell_{k, g}^{-i}\big)
       + \sum_{g \in \mathcal{A}(\svec_{2,i}; k_0)} \big(\ell_{k_0, g}^{-i} - \ell_{k_0, g}^{+i}\big).
\label{supp:eq:RZ2}
\end{align}

From a computational perspective, it is advantageous to express the likelihood in ratio form, as the terms unaffected by the proposed update---denoted $U$---need not be recomputed at each sampled location. This avoids redundant evaluation of group likelihoods that remain unchanged, thereby significantly reducing the per-iteration cost.

Moreover, when evaluating the full conditional distribution over all candidate labels $k = 1, \ldots, K$, the use of the likelihood ratio permits cancellation of the normalising constant across all choices of $k$. Specifically, when computing probabilities of the form
\begin{align*}
\Pr\big(Y_{1,i}=k \mid \Yvec_{1}^{\setminus i} , \Zvec_1, \Zvec_2,\Xvec, \thetab_1, \thetab_2, \hat{\beta} \big)
&\propto
p\big(\Zvec_2 \mid Y_{1,i}=k,\, \Yvec_{1}^{\setminus i},\, \Xvec, \thetab_2\big)\times \ldots \\
&\propto \exp\big(R_{\Zvec_2}(\svec_2; k_0, k)\big) \times \ldots,
\end{align*}
the normalising constant required to ensure probabilities sum to one over $k$ is independent of the proposed label. Consequently, only the numerator terms are needed to compare or rank proposals.

In addition to enabling efficient computation of conditional probabilities, this ratio-based formulation provides a convenient mechanism for incrementally updating the total log-likelihood whenever a change to $Y_1(\svec_{2,i})$ is accepted.

\subsection{Estimation of model parameters}
\label{supp:sec:gibbs-other}
Given the current configuration of the latent class field $\Yvec_1$, the remaining model parameters are updated according to the MCMC scheme described in Section~\ref{sec:fit} of the manuscript.

\section{Prediction of continuous geotechnical properties}
\label{supp:prediction}
We now describe how to generate predictions of the latent geotechnical property $Y_2(\cdot)$ over a finite set of unobserved locations.
Let $\mathcal{S}_{\p} \subset \mathcal{S}$ denote a set of prediction locations, typically chosen to form a depth profile, lateral cross-section, or 3D region of interest.
The goal is to obtain draws from the posterior predictive distribution $p\big( \Yvec_2^{(\p)} \mid \Zvec_1, \Zvec_2, \Xvec^{(\p)} \big),$
where $\Yvec_2^{(\p)} \equiv \big( Y_2(\svec_{\p,1}), \ldots, Y_2(\svec_{\p,N_\p}) \big)'$ for prediction locations $\svec_{\p,1},\ldots,\svec_{\p,N_\p} \in \mathcal{S}_\p$.

From the MCMC estimation process, we obtain $B$ posterior samples of the latent class field $Y_1(\cdot)$ over $\mathcal{S}_1$, denoted $
\{\Yvec_{1,b}\}_{b=1}^B,$ with $\Yvec_{1,b} \equiv (Y_{1,b,1},\ldots,Y_{1,b,N_1})'$, and corresponding parameter draws $\{\thetab_{2,b}\}_{b=1}^B$ for the class-specific Gaussian process model.

For each posterior draw $b = 1,\ldots,B$, we generate a posterior predictive draw $\Yvec_{2,b}^{(\p)}$ as follows:

\begin{enumerate}
    \item For each prediction location $\svec_{\p,i} \in \mathcal{S}_\p$, assign the class label via nearest-neighbour projection onto $\mathcal{S}_1$. That is, take $Y_1^{(b)}(\svec_{\p,i}) \equiv Y_{1,b,\,i^*}$, where  $i^* = \arg\min_{j \in \{1,\ldots,N_1\}} \|\svec_{\p,i} - \svec_{1,j}\|$. Let $\Yvec_{1,b}^{(\p)} \equiv \big( Y_1^{(b)}(\svec_{\p,1}), \ldots, Y_1^{(b)}(\svec_{\p,N_\p}) \big)'$ denote the resulting vector of predicted class labels.

    \item For each $k \in \{1,\ldots,K\}$, define the class-specific index set of prediction locations as $
    \mathcal{P}^{(b)}_k \equiv \big\{ i \in \{1,\ldots,N_\p\} : Y_1^{(b)}(\svec_{\p,i}) = k \big\}$, and let $\Yvec_{2,b}^{(k,\p)} \equiv \{ Y_2^{(b)}(\svec_{\p,i}) : i \in \mathcal{P}^{(b)}_k \}$ denote the corresponding predicted values for class $k$.

    \item   Given the current draw $\Yvec_{1,b}$ and GP parameters $\thetab_{2,b}$, compute the posterior predictive distribution of $\Yvec_{2,b}^{(k,\p)}$ conditional on observed measurements $\Zvec_2^{(k)}$ and covariates $\Xvec^{(k,\p)}$ at the locations in class $k$.

    This step involves forming the multivariate Gaussian predictive distribution:
    \[
    \Yvec_{2,b}^{(k,\p)} \,\big|\,\Yvec_1^{(k,\p)},  \Zvec_2^{(k,\p)}, \thetab_{2,b} \;\sim\; \mathcal{N}\big( \mub^{(k,\p)}_{b},\, \tilde\Sigmab^{(k,\p)}_{b} \big),
    \]
    where the mean and covariance are computed from the class-$k$ GP model, as described in Section~\ref{sec:model-Z2}.

    \item The predicted vector $\Yvec_{2,b}^{(\p)}$ is obtained by combining all class-specific predictions and reordering them to match the original order of the 3D lattice $\Ss_1$.
\end{enumerate}

\section{Parameter recovery in the simulation study}
\label{supp:sim-eresults}
\begin{figure}
    \centering
    \includegraphics[]{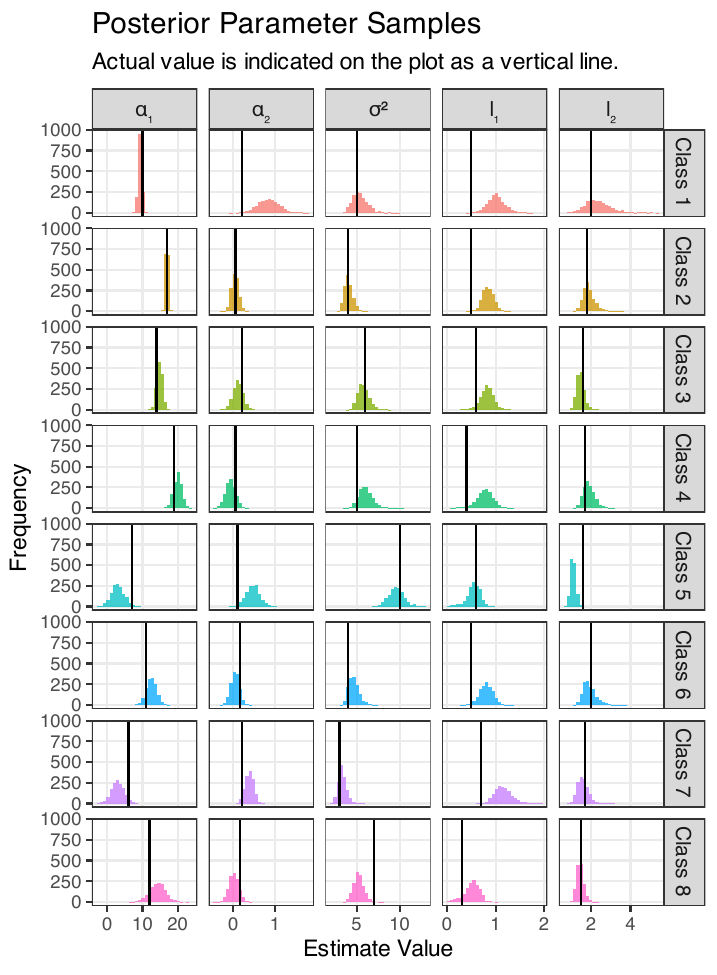}
    \caption{Posterior recovery of true parameter values in the simulation study. True values (vertical lines) are compared with posterior samples of class-specific regression coefficients ($\boldsymbol{\alpha}$), process variances ($\boldsymbol{\sigma}^2$), and spatial length scales ($\ell_1=\ell_L$, $\ell_2=\ell_D$) across eight simulated geological strata.}
    \label{supp:fig:syn_par}
\end{figure}

In this section we assess the model's ability to recover the true parameter values used in the simulation. Figure~\ref{supp:fig:syn_par} presents posterior samples for the class-specific regression coefficients ($\boldsymbol{\alpha}$), process variances ($\boldsymbol{\sigma}^2$), and length scales ($\boldsymbol{\ell}_L$, $\boldsymbol{\ell}_D$), denoted in the figure as $\ell_1$ and $\ell_2$, respectively. The posterior samples for the measurement error variance $\tau^2$ are not shown.

Across the eight classes, the intercept ($\alpha_1$) and slope ($\alpha_2$) parameters are well recovered, with posteriors tightly centred on the true values for most strata. Minor deviations are observed in the spatially dominant classes (e.g., class~1 and class~8), while a larger bias occurs for the less dominant class~5 stratum, where limited data support leads to broader posteriors. The process variances ($\sigma^2$) are similarly well estimated, though several classes display mild right skew, indicating residual uncertainty in the magnitude of spatial variability.

The lateral and vertical length scales ($\ell_1$ and $\ell_2$, corresponding to $\ell_L$ and $\ell_D$) are distinctly separated, capturing the true anisotropy with longer vertical correlation and shorter lateral dependence. Posterior uncertainty varies across classes, with flatter distributions in strata supported by fewer observations. The measurement error variance, while not shown in Figure~\ref{supp:fig:syn_par}, yields a posterior mean of $\hat{\tau}^2 \approx 0.85$, substantially exceeding the true value of $\tau^2 = 0.25$ (i.e., $\tau = 0.5$). This overestimation reflects the model's conservative attribution of unresolved local variation near class boundaries to unstructured noise rather than structured spatial effects. Overall, the reasonable agreement between the posterior and true parameters confirms that GeoMix effectively disentangles structured spatial effects from residual variability, demonstrating robust recovery of both regression and covariance components.

\section{MCMC diagnostics for the simulation study}
\label{supp:sim-mcmc}
We assessed convergence and mixing of the GeoMix MCMC algorithm using standard diagnostic tools, including trace plots, autocorrelation functions, and effective sample size (ESS) summaries. Four parallel chains were each run for 4,000 iterations with 500 burn-in iterations. Retaining every tenth post-burn-in iteration and combining chains yielded $1{,}400$ posterior samples. Diagnostics and predictions were conducted on these $1{,}400$ posterior samples.

\begin{figure}
\centering
\includegraphics[width=\textwidth]{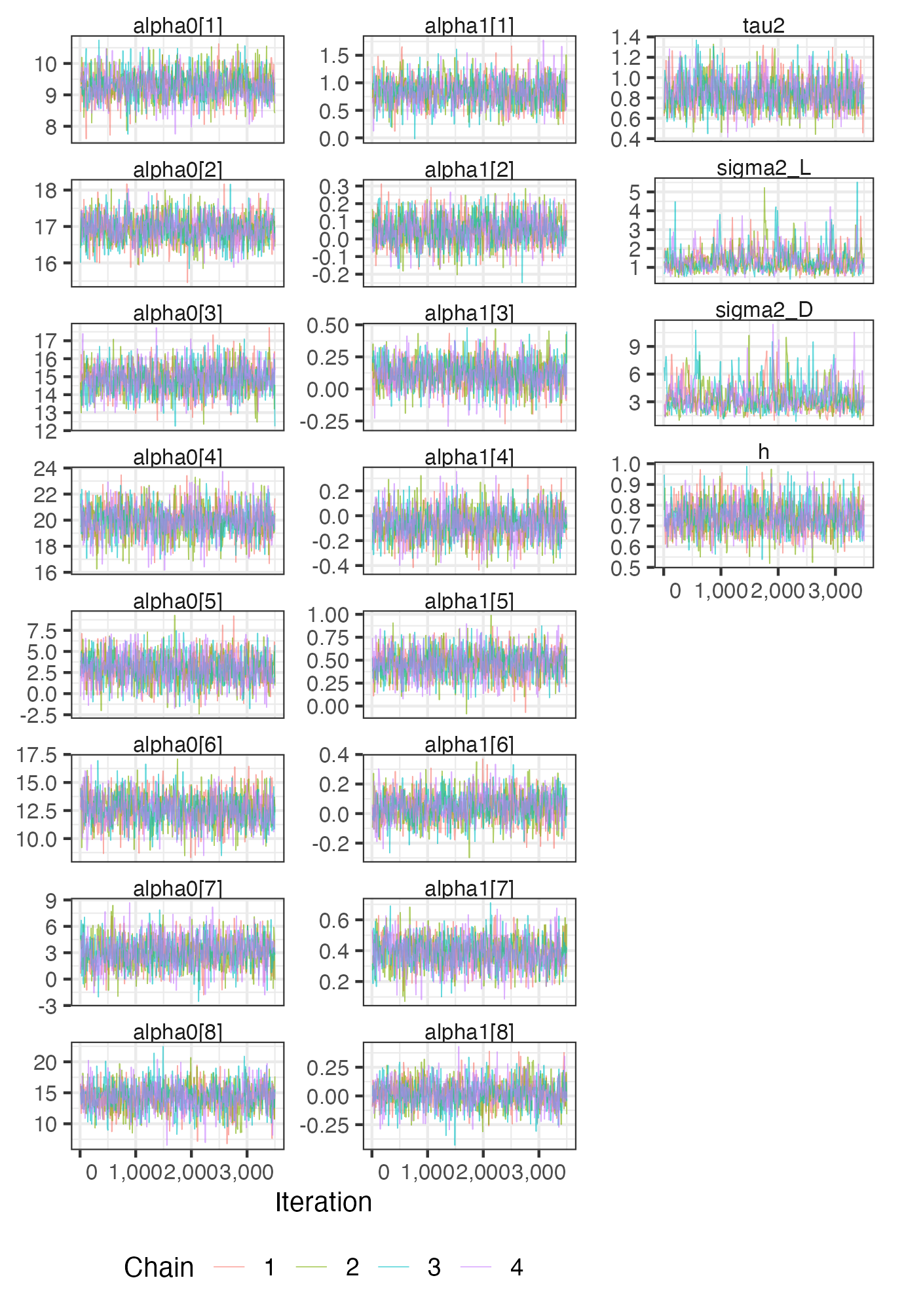}
\caption{Trace plots for the regression coefficients $(\alpha_{0,k}, \alpha_{1,k})$, $k=1,\ldots,8$, pooled length-scale variances $\sigma^2_L$ and $\sigma^2_D$, measurement noise variance $\tau^2$, and boundary bandwidth $h$ across the four MCMC chains. Every tenth iteration is shown for clarity.}
\label{supp:fig:trace_diag_syn1}
\end{figure}

\begin{figure}
\centering
\includegraphics[width=\textwidth]{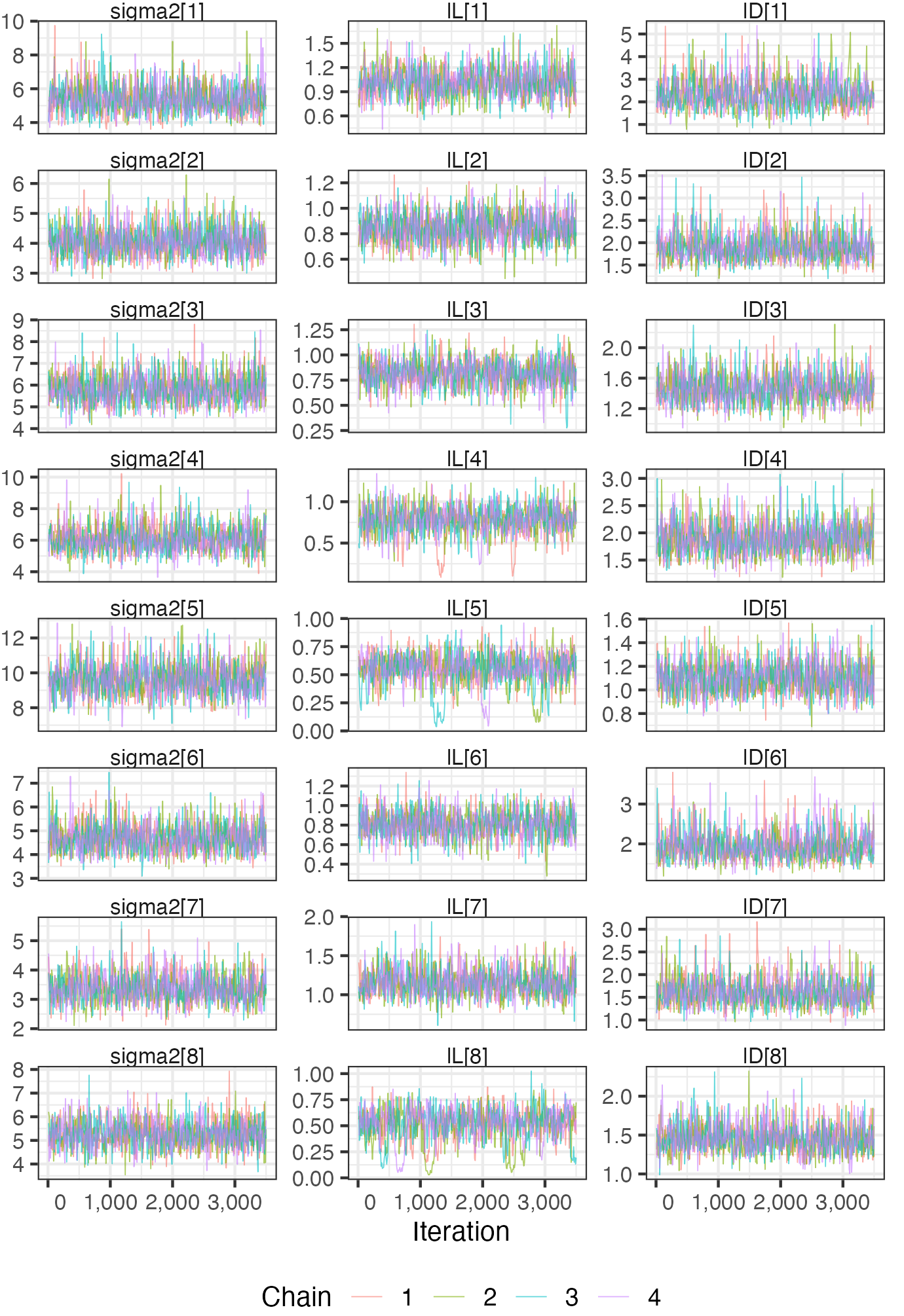}
\caption{Trace plots for the class-specific variance parameters $\sigma_k^2$ and spatial length-scales $(\ell_{L,k}, \ell_{D,k})$, $k=1,\ldots,8$, across the four MCMC chains. Every tenth iteration is shown for clarity.}
\label{supp:fig:trace_diag_syn2}
\end{figure}

\begin{figure}
\centering
\includegraphics[width=\textwidth]{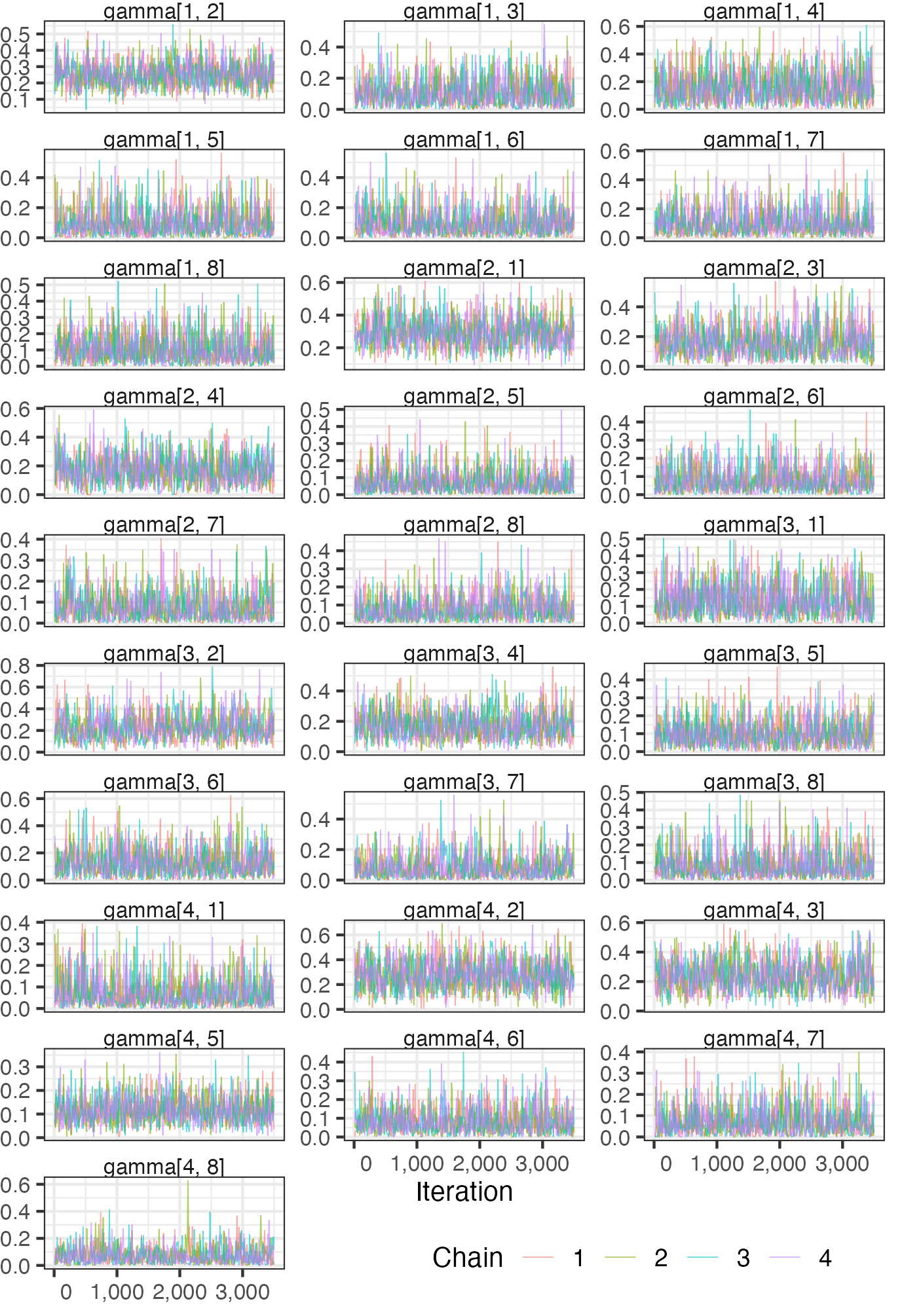}
\caption{Trace plots for the confusion matrix basis weights $\gamma_{k,\ell}$, $\ell=1,\ldots,8$, for $k=1,\ldots,4$, across the four MCMC chains. Every tenth iteration is shown for clarity.}
\label{supp:fig:trace_diag_syn3}
\end{figure}

\begin{figure}
\centering
\includegraphics[width=\textwidth]{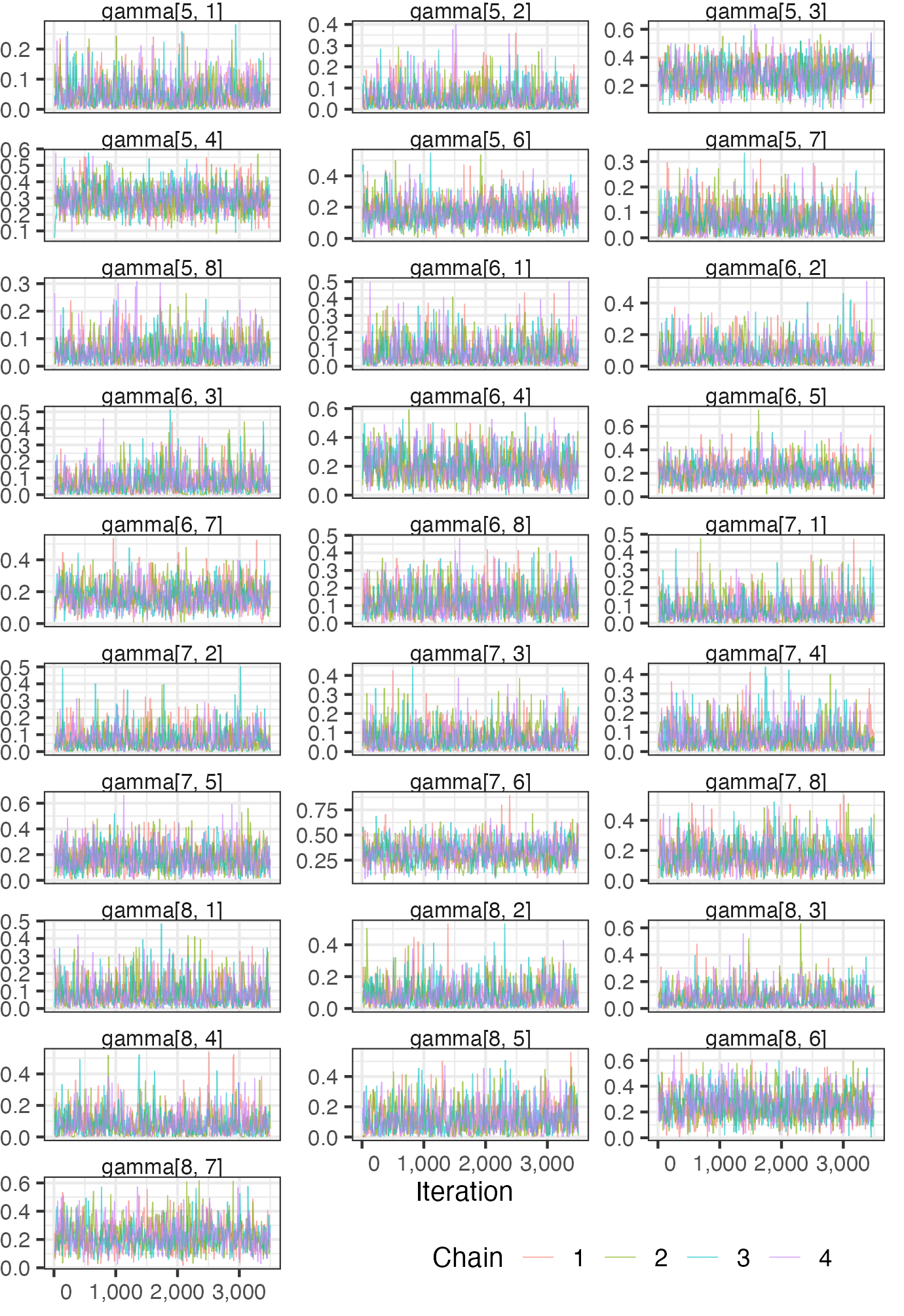}
\caption{Trace plots for the confusion matrix basis weights $\gamma_{k,\ell}$, $\ell=1,\ldots,8$, for $k=5,\ldots,8$, across the four MCMC chains. Every tenth iteration is shown for clarity.}
\label{supp:fig:trace_diag_syn4}
\end{figure}

Figures~\ref{supp:fig:trace_diag_syn1}--\ref{supp:fig:trace_diag_syn4} present trace plots for all model parameters across all chains, with every tenth iteration shown for clarity. These include the regression coefficients $(\alpha_{0,k}, \alpha_{1,k})$ for $k = 1,\ldots,8$, the pooled variance hyperparameters for the class-specific lateral and depth length-scales, $\sigma^2_L$ and $\sigma^2_D$, the measurement error variance $\tau^2$, and the boundary bandwidth $h$ (Figure~\ref{supp:fig:trace_diag_syn1}). Figure~\ref{supp:fig:trace_diag_syn2} shows the class-specific variance parameters $\sigma_k^2$ and spatial length-scales $(\ell_{L,k}, \ell_{D,k})$ for $k = 1,\ldots,8$. Figures~\ref{supp:fig:trace_diag_syn3} and \ref{supp:fig:trace_diag_syn4} display the confusion matrix basis function weights $\gamma_{k,\ell}$ for $\ell = 1,\ldots,8$, split across $k = 1,\ldots,4$ and $k = 5,\ldots,8$, respectively. A summary of diagnostic statistics by parameter class is provided in Table~\ref{supp:tab:ess_by_class}, computed from the posterior samples used for prediction after combining all four thinned chains into one sample.   

The chains exhibit stable behaviour with no evidence of long-term trends or non-stationarity. All parameters mix well and fluctuate around a common stationary region after burn-in, indicating effective exploration of the posterior distribution. This is supported by effective sample size and $\hat{R}$ diagnostics, with all parameters achieving relatively large bulk ESS and tail ESS values, and $\hat{R}$ values close to 1. In particular, the regression coefficients $\alphab$, confusion matrix basis weights $\gamma_{k,\ell}$, and bandwidth parameter $h$ exhibit rapid mixing, low autocorrelation, and large effective sample sizes.

Slightly slower mixing is observed for some covariance-related parameters, particularly the lateral length-scales $\ell_{L,k}$ and the variance hyperparameters $\sigma_L^2$ and $\sigma_D^2$, which exhibit comparatively lower effective sample sizes and mild autocorrelation. However, even the weakest-mixing parameters remain well within standard convergence thresholds. The latent stratigraphic field $\Yvec_1$ similarly exhibits stable behaviour across chains, with no indication of mode switching or divergent behaviour.

Taken together, the diagnostics indicate satisfactory convergence and adequate sampling efficiency across all parameters, supporting reliable posterior inference.
\begin{table}[t]
\centering
\caption{Diagnostic summaries by parameter class for simulation study.}
\label{supp:tab:ess_by_class}
\begin{tabular}{lcccccc}
\hline
Parameter & Count & \makecell{Med\\$\hat{R}$} & \makecell{Max\\$\hat{R}$} & \makecell{Med\\ESS} & \makecell{Min\\ESS} & \makecell{Max\\MCSE/SD} \\
\hline
Regression coefficients ($\alpha$) & 16 & 1.002 & 1.006 & 1375 & 1286 & 0.028 \\
Basis weights ($\gamma_{k,\ell}$) & 56 & 1.002 & 1.010 & 900 & 654 & 0.036 \\
Boundary bandwidth ($h$) & 1 & 1.002 & 1.002 & 1417 & 1417 & 0.027 \\
Length scales $\ell_D$ & 8 & 1.002 & 1.008 & 1180 & 825 & 0.035 \\
Length scales $\ell_L$ & 8 & 1.002 & 1.010 & 1110 & 312 & 0.065 \\
Marginal variance parameters ($\sigma_k^2$) & 8 & 1.000 & 1.005 & 1282 & 927 & 0.033 \\
Other variances ($\tau^2$, $\sigma^2_L$, $\sigma^2_D$) & 3 & 1.002 & 1.006 & 517 & 481 & 0.047 \\
\hline
\end{tabular}
\end{table}

\begin{figure}[t]
\centering
\includegraphics[width=\textwidth]{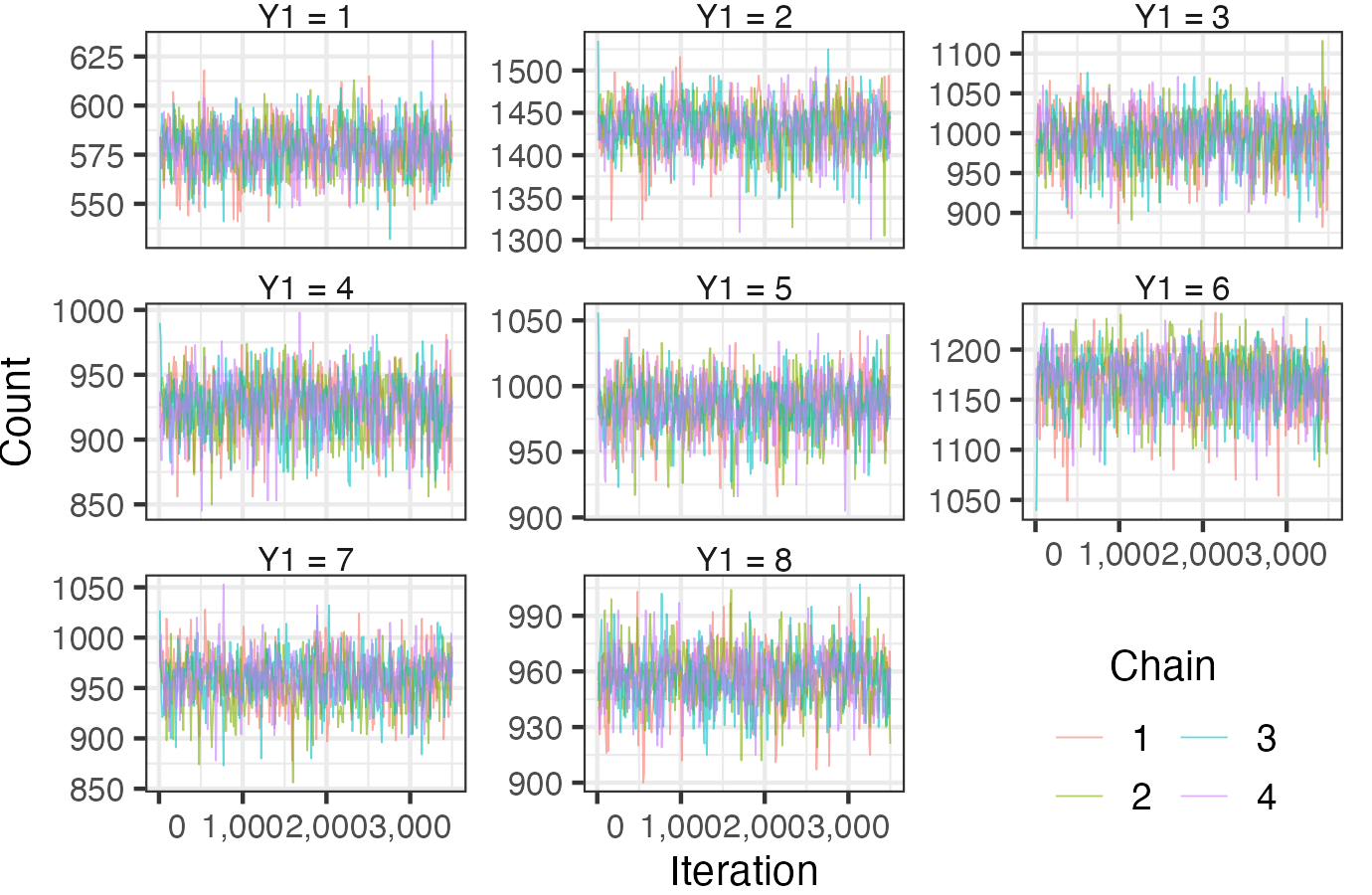}
\caption{Trace plots of the number of $\Ss_2$ locations assigned to each latent stratum $Y_1(\cdot)=k$, $k=1,\ldots,8$, across the four MCMC chains. Each panel corresponds to one stratum, with every tenth iteration shown. Counts are generally stable, indicating that overall stratigraphic composition is preserved during sampling, although strata $k=4$ and $k=6$ exhibit more pronounced variation, consistent with the regime-switching behaviour observed in covariance parameters.}
\label{supp:fig:trace_synY1}
\end{figure}
\section{MCMC diagnostics for the IJV wind farm zone application}
\label{supp:ijv-mcmc}

We assessed convergence and mixing of the GeoMix MCMC algorithm for the IJV--AB site using trace plots and standard summary diagnostics, following the approach in Section~\ref{supp:sim-mcmc}. Owing to the sparse lateral distribution of CPT observations relative to the spatial extent and resolution of the domain, the lateral length-scales $\ell_{L,k}$ were weakly identifiable and therefore fixed at their MAP estimates prior to MCMC. In this setting, lateral correlation ranges smaller than the spacing between nearby CPT locations produce covariance structures that are difficult to distinguish from measurement noise, leading to poor mixing.  

Four chains were run for $3{,}500$ post--burn-in iterations (following $500$ burn-in iterations), with every tenth iteration retained for posterior inference, yielding $1{,}400$ posterior samples across all chains.

\begin{figure}
\centering
\includegraphics[width=\textwidth]{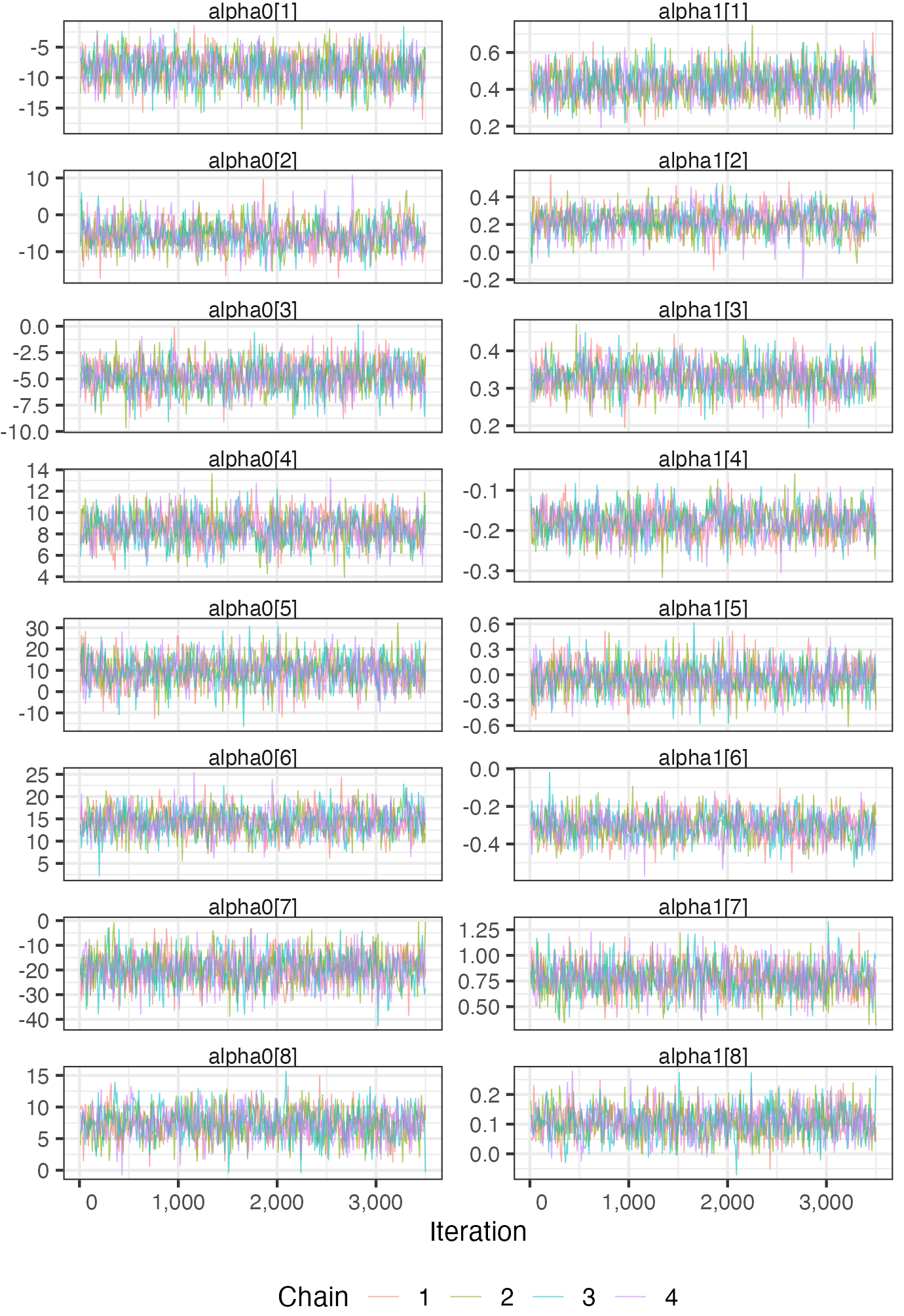}
\caption{Trace plots for the regression coefficients $(\alpha_{0,k}, \alpha_{1,k})$ for $k=1,\ldots,8$, across the four MCMC chains of the IJV-AB site data. For visual clarity, only every tenth iteration is plotted.}
\label{supp:fig:trace_diag_ijv1}
\end{figure}

\begin{figure}
\centering
\includegraphics[width=\textwidth]{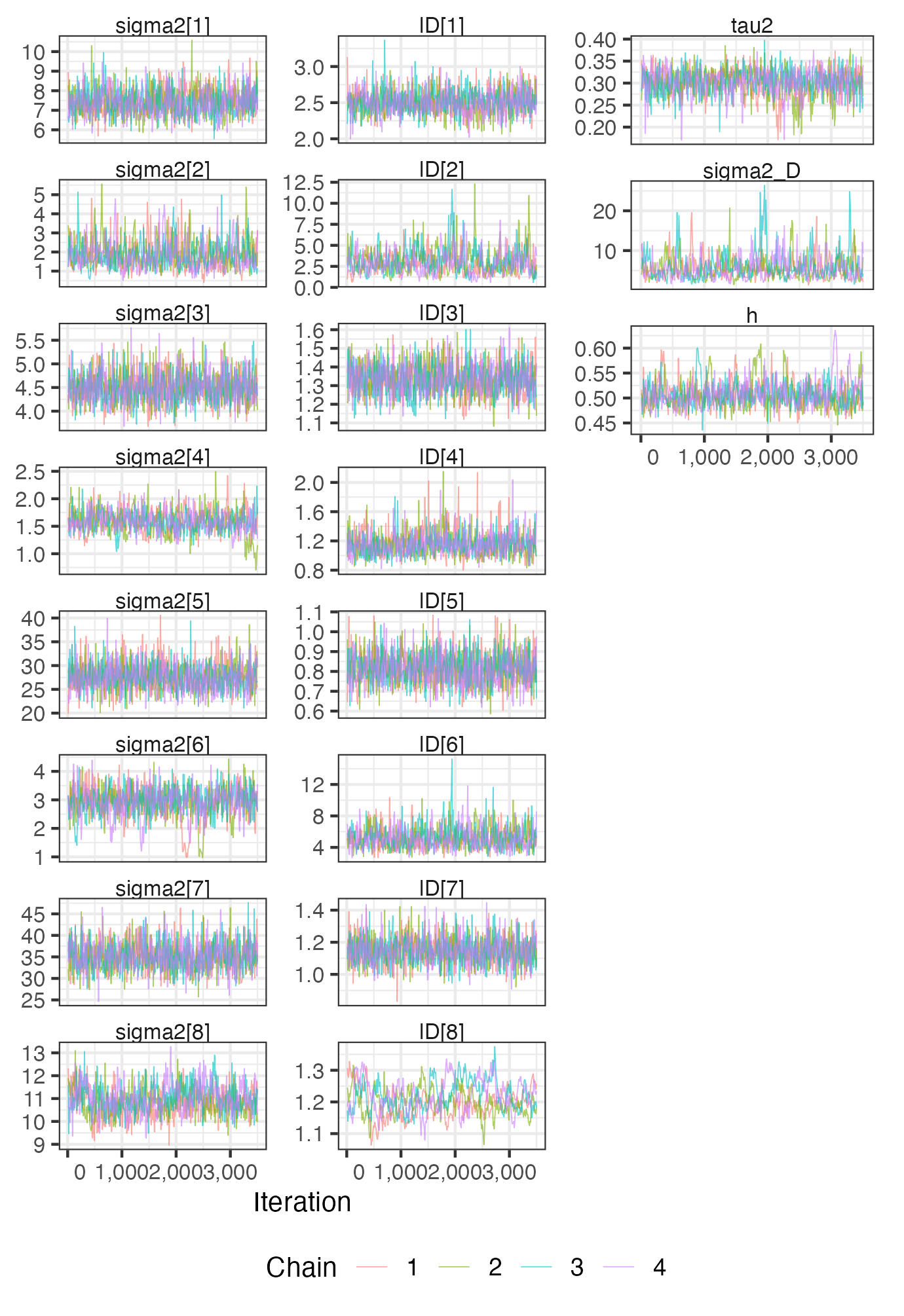}
\caption{Trace plots for the class-specific variance parameters $\sigma_k^2$ and vertical length-scales $\ell_{D,k}$, for $k=1,\ldots,8$, as well as the pooled length scale variances $\sigma^2_L$ and  $\sigma^2_D$, the measurement noise variance $\tau^2$, and the boundary bandwidth $h$, across the four MCMC chains of the IJV-AB site data. For visual clarity, only every tenth iteration is plotted.}
\label{supp:fig:trace_diag_ijv2}
\end{figure}

\begin{figure}
\centering
\includegraphics[width=\textwidth]{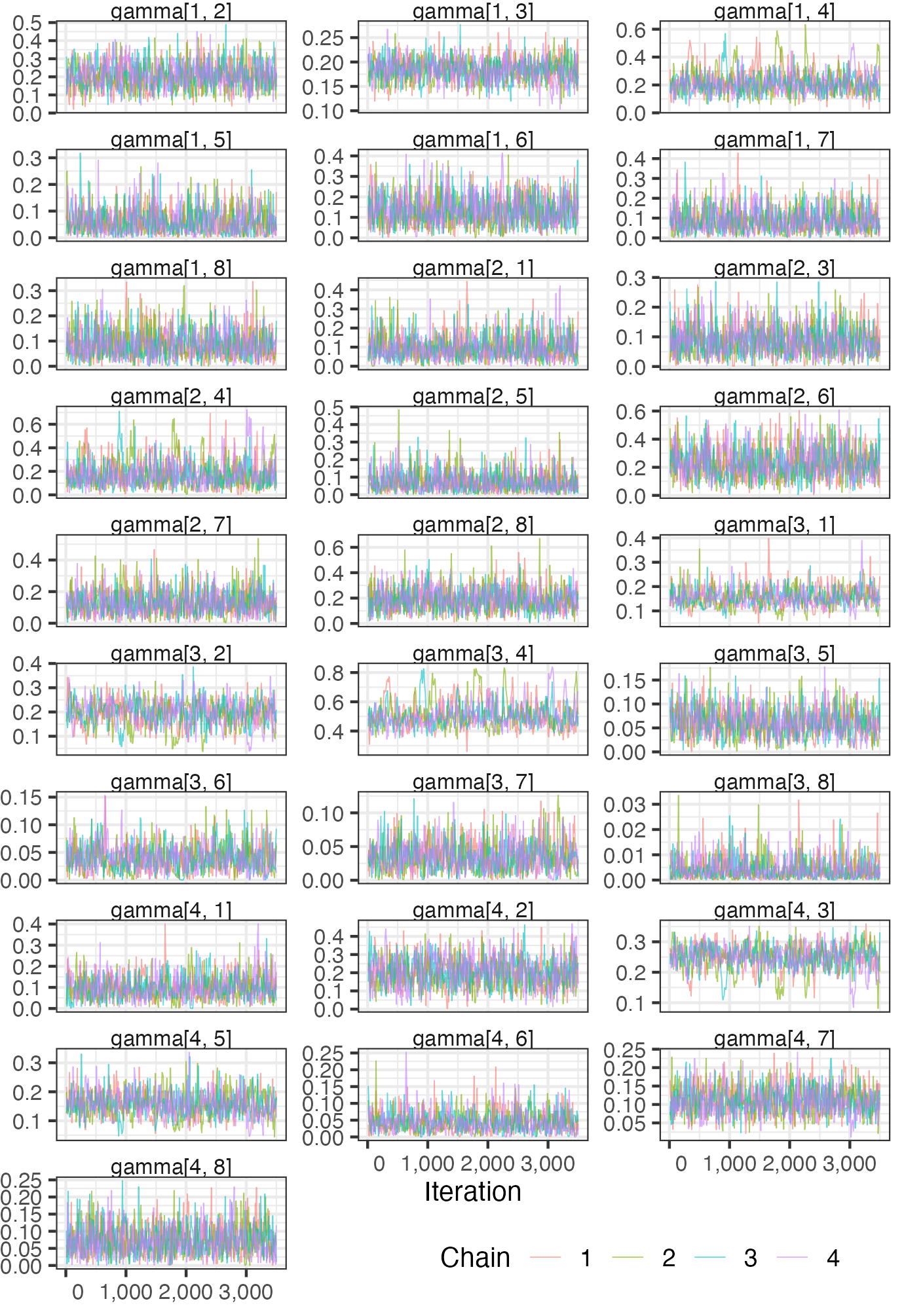}
\caption{Trace plots for the confusion matrix basis function weights $\gamma_{k,\ell}$ for $k=1,2,3,4$ and $\ell = 1,\ldots,8$ across the four MCMC chains of the IJV-AB site data. For visual clarity, only every tenth iteration is plotted.}
\label{supp:fig:trace_diag_ijv3}
\end{figure}

\begin{figure}
\centering
\includegraphics[width=\textwidth]{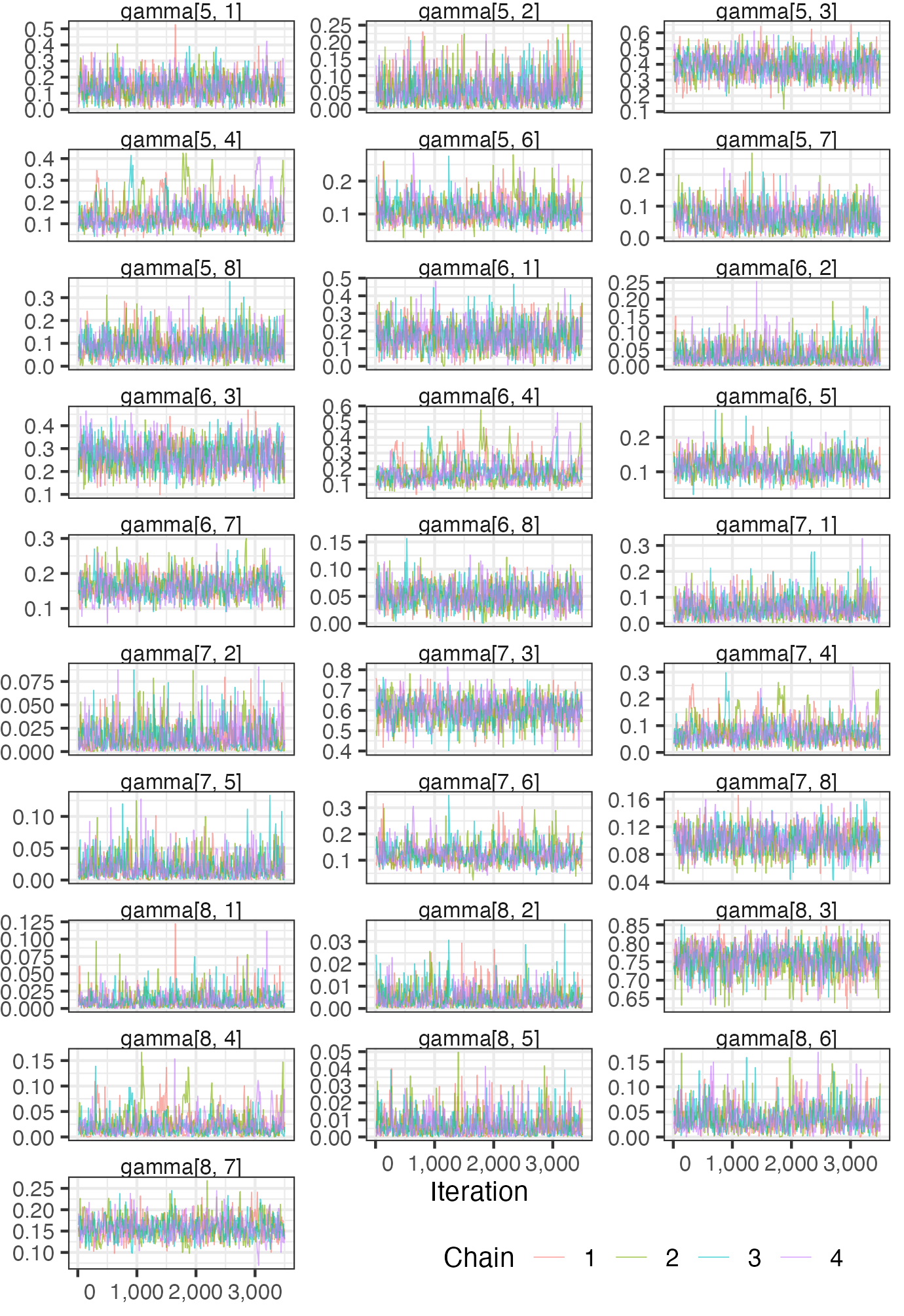}
\caption{Trace plots for the confusion matrix basis function weights $\gamma_{k,\ell}$ for $k=5,6,7,8$ and $\ell = 1,\ldots,8$ across the four MCMC chains of the IJV-AB site data. For visual clarity, only every tenth iteration is plotted.}
\label{supp:fig:trace_diag_ijv4}
\end{figure}

\begin{figure}[t]
\centering
\includegraphics[width=\textwidth]{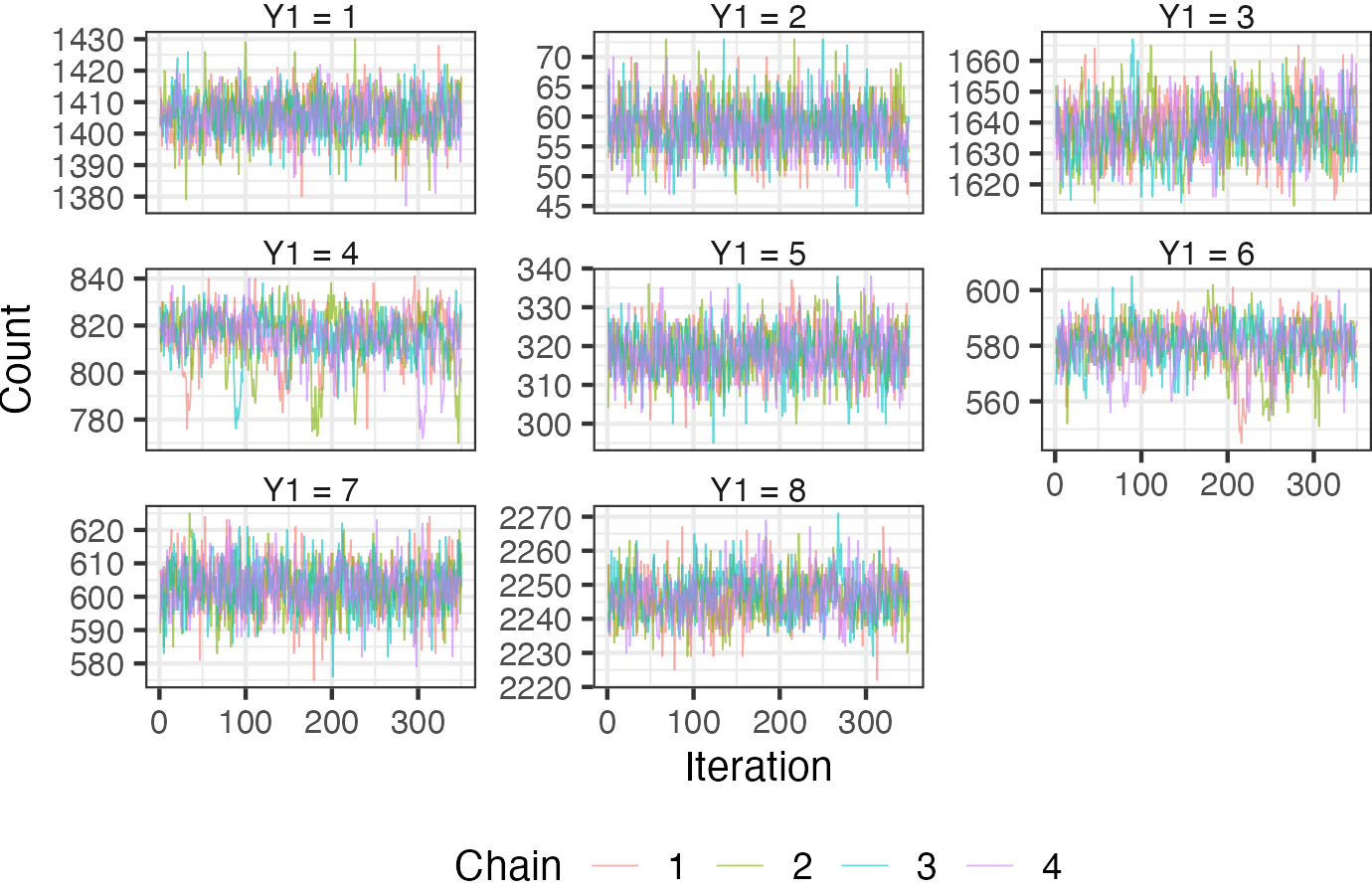}
\caption{Trace plots of the number of $\Ss_2$ locations (where CPT values are observed) assigned to each latent stratum, $Y_1(\cdot)=k$ for $k=1,\ldots,8$, across the four MCMC chains for the IJV-AB site data. Each panel corresponds to one stratum, with every tenth count shown over the full MCMC run for each chain. Overall, the counts are reasonably stable across iterations, indicating that the broad stratigraphic composition is largely preserved throughout sampling.}
\label{supp:fig:trace_Y1}
\end{figure}

Figures~\ref{supp:fig:trace_diag_ijv1}--\ref{supp:fig:trace_diag_ijv4} present trace plots for all model parameters across all chains, with every tenth iteration shown for clarity. Figure~\ref{supp:fig:trace_diag_ijv1} shows the regression coefficients $(\alpha_{0,k}, \alpha_{1,k})$ for $k=1,\ldots,8$. Figure~\ref{supp:fig:trace_diag_ijv2} displays the pooled variance hyperparameters $\sigma_L^2$ and $\sigma_D^2$, the measurement error variance $\tau^2$, the boundary bandwidth $h$, and the class-specific variance and depth length-scale parameters $(\sigma_k^2,\ell_{D,k})$ for $k=1,\ldots,8$. Figures~\ref{supp:fig:trace_diag_ijv3} and \ref{supp:fig:trace_diag_ijv4} display the confusion matrix basis function weights $\gamma_{k,\ell}$ for $\ell=1,\ldots,8$, split across $k=1,\ldots,4$ and $k=5,\ldots,8$, respectively. A summary of diagnostic statistics by parameter class is provided in Table~\ref{supp:tab:ijv_ess_by_class}, computed from the posterior samples retained for prediction. Figure~\ref{supp:fig:trace_Y1} summarises the evolution of the latent stratigraphic field through class occupancy counts. 

Overall, the chains exhibit stable behaviour with no evidence of long-term drift, and all parameters mix relatively well and fluctuate around a common stationary region. In particular, the regression coefficients $\alphab$ and confusion matrix basis function weights $\gamma_{k,\ell}$ exhibit good mixing, relatively low autocorrelation, and strong agreement across chains.

Mixing is somewhat slower for covariance-related parameters, particularly the depth length-scales $\ell_{D,k}$ and variance parameters $\sigma_k^2$, which exhibit higher autocorrelation and reduced effective sample sizes relative to the remaining parameters. The most notable case is $\ell_{D,8}$, which exhibits the slowest mixing and largest $\hat{R}$ across all parameters, while $\sigma_8^2$ also exhibits comparatively slow mixing. This behaviour is consistent with the geological interpretation of stratum $8$, which corresponds to the deepest and least clearly resolved unit in the ground model. In these deeper regions, seismic reflectors become less distinct and stratigraphic boundaries are more spatially irregular due to long-term geological deformation and heterogeneous sediment deposition, resulting in a stratum with weaker spatial identifiability and greater within-class heterogeneity. 

The latent stratigraphic field $\Yvec_1$ similarly exhibits moderate autocorrelation and slower mixing than the regression and misclassification parameters. However, as shown in Figure~\ref{supp:fig:trace_Y1}, the number of locations assigned to each stratum remains broadly stable across iterations, indicating that the large-scale stratigraphic structure is preserved throughout sampling.

Overall, the diagnostics indicate satisfactory convergence and adequate sampling efficiency for the model parameters, supporting adequate posterior inference despite slower mixing in a small subset of covariance parameters.

\begin{table}[h]
\centering
\caption{Diagnostic summaries by parameter class for the IJmuiden Ver wind farm zone application. Overall, parameters exhibit satisfactory diagnostics.}
\label{supp:tab:ijv_ess_by_class}
\begin{tabular}{lcccccc}
\hline
Parameter & Count & \makecell{Med\\$\hat{R}$} & \makecell{Max\\$\hat{R}$} & \makecell{Med\\ESS} & \makecell{Min\\ESS} & \makecell{Max\\MCSE/SD} \\
\hline
Regression coefficients ($\alpha$) & 16 & 1.001 & 1.004 & 1345 & 899 & 0.034 \\
Basis weights ($\gamma_{k,\ell}$) & 56 & 1.003 & 1.012 & 781 & 304 & 0.063 \\
Boundary bandwidth ($h$) & 1 & 1.006 & 1.006 & 431 & 431 & 0.055 \\
Length scales $\ell_D$ & 8 & 1.003 & 1.069 & 751 & 76 & 0.116 \\
Variance parameters ($\sigma_k^2$) & 8 & 1.008 & 1.034 & 551 & 169 & 0.077 \\
Other variances ($\tau^2$, $\sigma^2_D$) & 2 & 1.008 & 1.010 & 388 & 349 & 0.061 \\
\hline
\end{tabular}
\end{table}

\newpage